\documentclass[twocolumn,twocolappendix]{aastex701}

\usepackage{comment}
\newcommand{\nbodysix}{{\sc nbody6++gpu}}

\defcitealias{Gieles2023}{GG23}
\graphicspath{{./}{Figures/}}
\shorttitle{Reconstructing the GCIMF}
\shortauthors{E. Moreno-Hilario et al.}
\accepted{October 7, 2026}
\graphicspath{{./}{Figures/}}

\begin{document}

\title{Reconstructing the Globular Cluster Initial Mass Function from Present-day Globular Cluster Systems}

\author[orcid=0000-0002-6906-2379,gname=Elizabeth,sname=Moreno Hilario]{Elizabeth Moreno-Hilario}
\affiliation{Department of Astronomy, Tsinghua University, 30 Shuangqing Road, Beijing 100084, China}
\email{elimorh@tsinghua.edu.cn}
\email[show]{elimorh@tsinghua.edu.cn}

\author[orcid=0000-0002-1253-2763,gname=Hui,sname=Li]{Hui Li} 
\affiliation{Department of Astronomy, Tsinghua University, 30 Shuangqing Road, Beijing 100084, China}
\email{}

\author[orcid=0000-0002-5749-8255,gname=Luis,sname=Martinez-Medina]{Luis A. Martinez-Medina}
\affiliation{Instituto de Astronomía, Universidad Nacional Autónoma de México, A. P. 70-264, C.P. 04510, CDMX, México}
\email{}

\correspondingauthor{Elizabeth Moreno Hilario}

\begin{abstract}

The near-universal turnover mass of the present-day globular cluster mass function (GCMF), $M_{\rm TO} \sim 2 \times 10^5\ {\rm M_\odot}$, is a well established observational feature across galaxies of different types and masses, providing an important empirical benchmark for understanding the globular cluster initial mass function (GCIMF). Competing explanations of this property invoke either dynamical evolution from an initial power-law distribution or an imprint of cluster formation physics. We address this problem by reconstructing the high-mass regime of the GCIMF by inverting the mass loss due to dynamical evolution of present-day globular cluster systems across a wide range of host galaxy masses. Our method is based on a host-galaxy-dependent mass-loss model calibrated by direct $N$-body simulations in time-dependent tidal fields, enabling a mapping between observed cluster masses and their progenitor values without assuming a priori a functional form for the GCIMF. We apply our method to galaxies spanning halo masses of $\sim10^{9}$ - $10^{12}\ {\rm M_\odot}$, combining systems with individually measured globular cluster masses as well as large statistical samples constructed from observed global properties. The recovered GCIMFs are systematically shifted towards higher masses and exhibit a power-law behavior at the high-mass end. The inferred slopes vary across galaxies and show a strong correlation with host halo mass, with more massive galaxies exhibiting steeper high-mass slopes. Our results suggest that the slope of the GCIMF depends on the galactic properties and provides a direct empirical link between present-day globular cluster systems and their high-redshift progenitors.

\end{abstract}



\section{Introduction}
\label{sec:Introduction}
Globular clusters (GCs) are among the oldest stellar systems in the Universe and provide valuable constraints on the early stages of galaxy formation and assembly. A large fraction of present-day GCs are metal-poor and have ages of $\gtrsim12$\,Gyr, indicating that they formed at high redshift, likely around the epoch of reionization \citep{Forbes2018}. Because most GCs have survived for more than $10$\,Gyr, their present-day properties encode information about both their formation conditions and their subsequent dynamical evolution within their host galaxies \citep{Brodie_Strader2006}. Direct observations of GC formation sites remain extremely challenging due to their small sizes and large distances. However, recent observations with the James Webb Space Telescope (JWST) have begun to reveal candidate proto–globular clusters at redshifts up to $z\sim10$ in strongly lensed systems \citep[e.g.][]{Mowla2022,Vanzella2023,Adamo2024}. At the same time, increasingly high-resolution hydrodynamical simulations are beginning to resolve the formation of bound star clusters in cosmological environments \citep[e.g.][]{Kimm2016, Rieder2022, Garcia2023, Andersson2024}. Despite these advances, the physical conditions under which GCs formed remain an open question.

One of the most remarkable observational properties of GC systems is the near-universal shape of their mass and luminosity distributions \citep{Vesperini_2001}. The present-day globular cluster luminosity function (GCLF), often expressed in terms of cluster mass as the globular cluster mass function (GCMF), exhibits a characteristic peaked or log-normal shape with a turnover mass near $M_{\rm TO}\sim2\times10^5\,{\rm M_\odot}$ \citep{Jordan2007, Harris_2015}. Remarkably, this turnover mass varies only weakly across galaxies spanning several orders of magnitude in stellar mass and a wide range of morphologies and environments. The apparent universality of this feature has made the GCLF a useful distance indicator \citep{Rejkuba_2012}, but it also raises a fundamental challenge for models of GC formation and evolution.

A long-standing problem is understanding the relationship between the present-day GCMF and the globular cluster initial mass function (GCIMF). Young massive cluster (YMC) populations observed in nearby star-forming galaxies typically follow a power-law mass function with slope $dN/dM \propto M^{-2}$ \citep[e.g.][]{Zhang_1999, McCrady_2007}. Because YMCs represent the closest observable analogues to young GCs, it is often assumed that GCs formed with a similar initial mass distribution. If this is the case, the present-day peaked GCMF must arise primarily from dynamical evolution processes that preferentially destroy low-mass clusters over cosmic time.

Early analytical studies demonstrated that dynamical processes such as two-body relaxation, tidal evaporation, and gravitational shocks in a galactic potential can transform an initially featureless or power-law cluster mass distribution into a peaked luminosity function over a Hubble time \citep{McLaughlin_1994,Fall2001}. In this framework, the turnover mass corresponds to the characteristic mass scale at which clusters can survive for $\sim10$ - $12$\,Gyr in typical galactic tidal fields. Numerical studies further showed that cluster disruption rates depend sensitively on environmental properties such as the strength of the galactic tidal field and the orbital histories of clusters \citep{Gnedin_1997}. More recent theoretical models have emphasized the strong dependence of cluster formation and survival on the evolving galactic environment, including the density of the interstellar medium and the dynamical history of the host galaxy \citep{kruijssen15}.

Other studies have explored the possibility that the turnover in the GCMF may instead reflect the initial conditions of cluster formation. For example, \citet{McLaughlin_1996} investigated the formation of GC systems within protogalactic environments and suggested that the observed mass distribution could arise from a combination of formation physics and subsequent dynamical disruption. Similarly, \citet{Parmentier_2007} showed that rapid gas expulsion from embedded protoclusters within a power-law molecular cloud mass spectrum can produce a bell-shaped cluster mass function at early times, even before long-term dynamical evolution becomes dominant. In this scenario, the characteristic mass scale of the GCMF would be largely imprinted during the formation phase rather than generated by secular dynamical processes.

Distinguishing between these scenarios remains challenging because direct observational constraints on the GCIMF are only now beginning to emerge. Recent JWST observations have identified candidate proto-globular clusters at high redshift \citep[e.g.,][]{Vanzella2023,Adamo2024}, providing the first constraints on the high-redshift cluster population. More recently, \citet{Claeyssens2026} combined observations of star clusters in strongly lensed galaxies to derive an average cluster mass function at high redshift. However, current observations remain limited to relatively small samples and primarily probe the most massive clusters, preventing a robust determination of the full GCIMF.

Most GCs formed at high redshift and have experienced substantial mass loss over their lifetimes, making it necessary to infer their progenitor populations through models of cluster evolution. In cosmological simulations and semi-analytic models, cluster mass loss is typically implemented using prescriptions in which evaporation driven by two-body relaxation and tidal shocks depends on cluster mass and the local tidal field strength \citep[e.g.][]{kruijssen15, Pfeffer2018, Chen2023, Gieles2023}. These prescriptions are usually calibrated against direct $N$-body simulations of star clusters. However, such simulations often adopt simplified setups in which clusters evolve on circular or eccentric orbits within fixed galactic potentials. While these controlled experiments provide valuable insight into cluster disruption processes, their idealized nature limits the accuracy with which these prescriptions can be extrapolated to the complex, time-varying environments of cosmological galaxy formation.

In general, these approaches demonstrate that dynamical mass-loss processes are capable of transforming an initial power-law or Schechter-like GCIMF into a peaked present-day GCMF. However, the details of this evolution depend sensitively on the adopted prescriptions for cluster formation and disruption. Several analytical and numerical studies have successfully reproduced the overall shape of the observed GCMF by including the effects of two-body relaxation, tidal disruption, and hierarchical galaxy assembly \citep{Fall2001,Vesperini_2001,kruijssen15,Choksi2018}. However, simultaneously reproducing the observed turnover mass scale within a cosmological framework remains an open problem. Many cosmological and semi-analytic models predict turnover masses that are significantly lower than the observed value of $\sim2\times10^5\ {\rm M_\odot}$, often by about an order of magnitude \citep[e.g.,][]{Pfeffer2018,Li2019,Rodriguez2023}. This discrepancy suggests that either the physical processes governing cluster disruption are not yet fully captured in current models, or that the GCIMF itself may differ from the commonly assumed universal power-law.

Most studies adopting this dynamical evolution framework implicitly assume that the GCIMF is universal across galaxies. In this picture, a common power-law initial mass distribution evolves through dynamical disruption to produce the near-universal turnover observed in present-day GC systems \citep{Fall2001}. However, this assumption remains largely untested observationally. An alternative possibility is that the GCIMF itself varies with the properties of the host galaxy and its formation environment. Because the efficiency of cluster formation, the strength of the tidal field, and the early dynamical evolution of clusters all depend on galaxy mass and structure, the resulting initial cluster population may not be universal.

In this work we test this possibility by reconstructing the GCIMF from present-day GC systems across a wide range of host galaxy masses. In \citet{Moreno-Hilario2024} (hereafter Paper I), we performed a suite of direct $N$-body simulations of GC evolution in external tidal fields representative of dwarf galaxies. In that study, five dwarf galaxies were selected from a high-resolution cosmological simulation, spanning halo masses of $M_{\rm h} \sim 10^{9}$ - $10^{11}\ {\rm M_\odot}$. The time dependent tidal histories experienced by clusters along their orbits were extracted from the simulation and coupled to direct $N$-body integrations using {\nbodysix}~\citep{Wang2015}. These simulations followed the long-term evolution of clusters spanning a range of initial masses, allowing us to quantify how two-body relaxation and tidal stripping drive cluster mass loss in realistic galactic environments. From this suite we derived a parameterized description of cluster mass-loss rates as a function of cluster mass and host galaxy mass. This calibration provides a physically motivated mapping between the present-day mass of a cluster and its progenitor mass, enabling empirical reconstruction of GC populations.

Using this methodology, we invert the mass evolution of present-day GC systems to estimate their progenitor mass distributions. For this purpose we compile a large sample of galaxies and their GC systems from the literature. For galaxies where individual GC masses are available, such as the Milky Way (MW) and nearby systems, we apply the mass-loss inversion cluster by cluster. For galaxies where only global GC system properties are known, we generate synthetic present-day mass distributions consistent with observational constraints and apply the same reconstruction statistically. By comparing the reconstructed GCIMFs across galaxy types ranging from dwarfs to MW-like hosts, we investigate whether the commonly assumed universal power-law GCIMF is consistent with the observed properties of GC systems, or whether the GCIMF itself varies systematically with galaxy properties.

This paper is organized as follows. In Section \ref{sec:Data} we describe the observational data and literature catalogs used to construct present-day GCMFs. Section \ref{sec:Methodology} presents the reconstruction procedure, in which present-day GC mass distributions are inverted using a host-galaxy-dependent mass-loss model calibrated on direct $N$-body simulations of star clusters. In Section \ref{sec:Results} we present representative reconstructed GCIMFs and characterize the dependence of the high-mass slope $\beta$ on host galaxy mass. Finally, Section \ref{sec:DiscussionAndConclusions} discusses the implications for the GCIMF problem and summarizes our conclusions.

\section{Observational Data and Present-Day Mass Distributions}
\label{sec:Data}

In order to recover the GCIMF in a given galaxy we require the mass of the galaxy and its present-day GCMF.
Hence, to obtain good statistics and cover a wide range of galaxy types and masses, we compile a large sample of GC systems from different public sources. The resulting sample combines galaxies with individually measured GC masses, photometric GC catalogs from the ACS Virgo Cluster Survey, and large statistical catalogs for which only integrated properties of the GC system are available. Together, these datasets allow us to construct present-day GC mass distributions spanning dwarf galaxies, low-mass spirals, and massive early-type systems.

\subsection{Galaxies with Individual GC Mass Measurements}
\label{subsec:IndividualGCMasses}
For several nearby galaxies, individual GC masses are available from the literature, enabling the construction of empirical present-day GCMFs.

\begin{figure}
    \centering
    \includegraphics[width=\columnwidth]{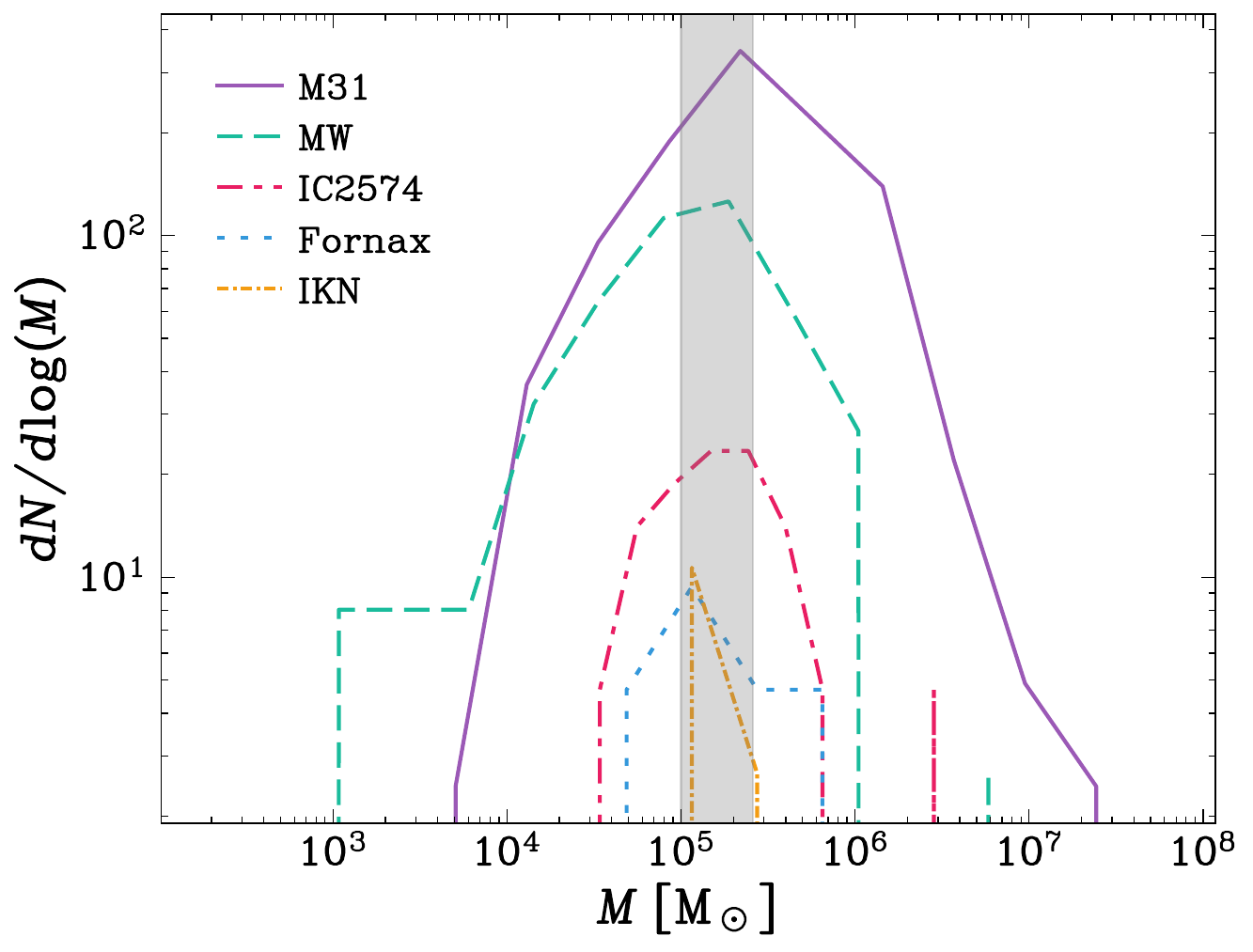}
    \caption{Present-day GCMFs of galaxies with individual GC mass measurements. The gray shaded region marks the expected turnover mass range $1 \times 10^5\ {\rm M_\odot} \leq M_{\rm TO} \leq 2.6 \times 10^5\ {\rm M_\odot}$ for all the galaxies studied in \citet{Jordan2007}. Despite the wide range of host galaxy masses, the distributions exhibit the characteristic near-universal log-normal form.}
    \label{fig:GCMF_present}
\end{figure}

Starting with the MW, we adopt dynamical GC masses from the Galactic Globular Cluster Database \citep{BaumgardtDatabase2023}, based on direct $N$-body modeling of velocity-dispersion profiles and surface-brightness distributions \citep{Baumgardt_2018}. Our sample includes 167 confirmed MW GCs with available dynamical mass estimates. The Galactic halo mass is taken to be $M_{\rm h} = 1.3 \times 10^{12}\,{\rm M_\odot}$ \citep{Posti_2019}. 

For M31, individual GC masses are taken from \citet{Caldwell_2016}, the sample includes 436 GCs. The halo mass $M_{\rm h} = 2.1 \times 10^{12}\ {\rm M_\odot}$ is adopted from \citet{Fardal_2013}, who derive a dynamical mass model based on kinematic and structural constraints of the Andromeda system.

For the low-mass spiral galaxy IC2574, we adopt GC masses compiled in the Local Volume Database \citep{Pace_2024}\footnote{The Local Volume Database is publicly available at \url{https://github.com/apace7/local_volume_database}.}. The halo mass is adopted from the recent dynamical analysis of \citet{Karim_2024}, yielding $M_{\rm h} = 8.54 \times 10^{10}\ {\rm M_\odot}$. This galaxy hosts 23 confirmed GCs and provides an important low-mass comparison system extending our sample toward halo masses similar to those explored in Paper I.

For the dwarf galaxy IKN, GC masses are taken from \citet{Tudorica_2015}. In the absence of a direct halo mass measurement, we estimate the halo mass using the empirical ratio between total GC system mass and halo mass,
\begin{equation}
\eta \equiv \frac{M_{\rm GCS}}{M_{\rm h}} = 3.5\times10^{-5},
\label{eq:eta_ratio}
\end{equation}
following the scaling relations reported by \citet{Harris_2015}, which gives $M_{\rm h} = 2.32 \times 10^{10}\ {\rm M_\odot}$. The IKN system contains five GCs.

For the Fornax dwarf spheroidal galaxy, GC masses are taken from \citet{deBoer_2016}. As for IKN, the halo mass is estimated using the same empirical $\eta$ relation, yielding $M_{\rm h} = 2.73 \times 10^{10}\ {\rm M_\odot}$. The Fornax system also contains five GCs.

Figure~\ref{fig:GCMF_present} shows the resulting present-day GCMFs for galaxies in our sample. Despite spanning more than four orders of magnitude in halo mass and diverse galactic environments from dwarfs to massive spirals, the distributions exhibit the characteristic peaked form with a turnover mass near $M_{\rm TO} \sim 2\times10^{5}\ {\rm M_\odot}$, consistent with previous studies of the near-universal present-day GCMF \citep[e.g.,][]{Jordan2007, Harris_2015, Forbes2018}. The observed cluster masses in these systems are used directly as inputs to the reconstruction procedure described in Section~\ref{sec:Methodology}.

\subsection{GC Masses from the Photometric Catalog of the ACS Virgo Cluster Survey}
\label{subsec:ACSVCS}
To extend our sample with a homogeneous set of GC mass estimates in dense environments, we make use of the ACS Virgo Cluster Survey \citep[ACSVCS;][]{Jordan2009}. The survey provides photometric measurements of GC candidates for early-type galaxies in the Virgo Cluster. Our final ACSVCS sample contains 89 galaxies with reconstructed GC systems.

The ACSVCS catalogs include positions, apparent magnitudes in the $g_{475}$ and $z_{850}$ bands, structural parameters such as half-light radii, and probabilities of cluster membership ($p_{\rm GC}$) derived from mixture modeling. We adopt the GC catalog of \citet{Jordan2009} and \citet{Peng2008}, selecting objects with $p_{\rm GC} > 0.5$ as probable GCs, following the original catalog definition. This threshold is used for consistency with the ACSVCS selection and does not significantly affect the resulting GC samples.

Individual GC masses are not provided in the catalog and are therefore derived from the photometry. For the selected clusters, we compute stellar masses using the color-dependent mass-to-light ratio from \citet{Bell2003}:
\begin{equation}
\log(M/L_z) = 0.322\,(g - z) - 0.171.
\label{eq:ML_relation}
\end{equation}

The resulting GC masses provide an internally consistent set of present-day GCMFs for Virgo Cluster galaxies, complementing the dynamically measured masses described in Section~\ref{subsec:IndividualGCMasses}. While these estimates rely on stellar population assumptions and are therefore subject to systematic uncertainties in the mass-to-light ratio, they enable us to include a significantly larger number of GC systems in cluster environments.

These photometrically derived GC masses are incorporated into the reconstruction pipeline in the same way as directly measured masses.

\subsection{Large Statistical Galaxy Samples}
\label{subsec:CatalogsData}
For large galaxy samples where individual GC masses are not available, we reconstruct the present-day GCMF stochastically using published measurements of global GC system properties. These include the total number of clusters $N_{\rm GC}$, the total GC system mass $M_{\rm GCS}$, and host galaxy stellar masses.

Our primary dataset is the compilation of \citet{Harris2013}, which provides a homogeneous catalog of GC system properties for several hundred galaxies spanning a broad range of morphological types and environments. The catalog includes measurements of $N_{\rm GC}$, estimates of $M_{\rm GCS}$, and host galaxy stellar masses compiled from the literature. To extend the analysis into the dwarf galaxy regime, we incorporate the catalog of \citet{Dornan_2026}, which compiles GC system measurements for nearby dwarf galaxies in groups and clusters. This dataset includes galaxies with stellar masses as low as $M_* \sim10^{6}$ - $10^{7}\ {\rm M_\odot}$ and provides estimates of $N_{\rm GC}$ together with host galaxy structural and stellar mass measurements. Some galaxies included in both compilations \citet{Harris2013} and \citet{Dornan_2026} originate from the ACS Virgo Cluster Survey. These systems are removed from the statistical catalogs in our analysis because they are treated separately using the original ACSVCS photometric catalogs, from which we derive individual GC masses directly (see Subsection~\ref{subsec:ACSVCS}).

To focus our analysis on the mass range where both observations and theoretical modeling are robust, we retain galaxies with stellar mass $M_* \lesssim 6\times10^{10}\ {\rm M_\odot}$. This selection ensures that we include galaxies with well-characterized individual GC system masses (see Subsection~\ref{subsec:IndividualGCMasses}) and aligns with our N-body simulations, which were performed in dwarf galaxy environments, so this selected range does not extend far beyond the dwarf regime. After removing overlapping systems and the galaxies treated separately through the ACSVCS analysis, the final statistical sample contains 352 galaxies. 

For each galaxy, we generate a synthetic present-day GC population by sampling $N_{\rm GC}$ clusters from a log-normal mass distribution with dispersion $\sigma_{\log M}=0.5$ dex. The turnover mass, $M_{\rm TO}$, is set according to the host galaxy stellar mass: for galaxies with $M_* > 10^9~{\rm M_\odot}$ we adopt $M_{\rm TO} = 2 \times 10^5\ {\rm M_\odot}$, while for lower-mass dwarf galaxies ($M_* \leq 10^9\ {\rm M_\odot}$) we adopt $M_{\rm TO} = 1.6 \times 10^5\ {\rm M_\odot}$, following the trend reported by \citet{Jordan2007}, who found that the turnover is slightly smaller in dwarfs compared to more massive galaxies. These values are consistent with other observational studies of the near-universal present-day GCMF \citep[e.g.,][]{Harris_2015, Forbes2018}. This stochastic reconstruction assumes that the present-day GCMF is well described by a log-normal distribution with only weak dependence on host galaxy mass.

The resulting synthetic mass distributions are then used as inputs to the reconstruction procedure described in Section~\ref{sec:Methodology}, where the dynamical inversion is applied cluster by cluster.

\subsection{Synthetic Present-Day Systems for Simulated Dwarfs}

In addition to the observational samples described above, we construct synthetic present-day GC systems for the dwarf galaxies analyzed in Paper I. These galaxies were selected from a high-resolution cosmological simulation and span halo masses of $\sim10^{9}$ - $10^{11}\ {\rm M_\odot}$ and stellar masses of $\sim5\times10^{6}$ - $5\times10^{8}\ {\rm M_\odot}$. The sample consists of five dwarf galaxies and provides tidal environments typical of low-mass systems. The main properties of these galaxies, including halo mass, initial GC system mass, and the mass-loss parameters used in our modeling, are summarized in Table\ref{tab:DwarfsPaperI}.
\begin{deluxetable}{lcccc}
\tablewidth{0pt}
\tablecaption{Properties of simulated dwarf galaxies A--E. \label{tab:DwarfsPaperI}}
\tablehead{
\colhead{Galaxy} &
\colhead{$M_{\mathrm{h}}$} & \colhead{$M_{\mathrm{GCS},0}$} & \colhead{$A$} & \colhead{$\alpha$} \\
\colhead{} & \colhead{$(M_\odot)$} & \colhead{$(M_\odot)$} & \colhead{} & \colhead{}
}
\startdata
A & $3.606 \times 10^{9}$  & $1.696 \times 10^{5}$  & 2.7315   & -0.4190 \\
B & $4.147 \times 10^{9}$  & $1.944 \times 10^{5}$  & 2.7805   & -0.4189 \\
C & $3.799 \times 10^{10}$ & $1.704 \times 10^{6}$  & 28.6816  & -0.5701 \\
D & $6.901 \times 10^{10}$ & $3.058 \times 10^{6}$  & 69.6092  & -0.6177 \\
E & $1.047 \times 10^{11}$ & $4.603 \times 10^{6}$  & 127.8145 & -0.6457 \\
\enddata
\tablecomments{Halo mass ($M_{\mathrm{h}}$) and initial globular cluster system mass ($M_{\mathrm{GCS},0}$). The dimensionless parameters $A$ and $\alpha$ describe the mass-loss model calibrated from $N$-body simulations.}
\end{deluxetable}

To assign present-day GC populations to these galaxies, we use the empirical scaling relation between GC system mass and host galaxy halo mass. Following the definition of $\eta$ introduced in Equation~\ref{eq:eta_ratio}, we estimate the total GC system mass as
\begin{equation}
M_{\mathrm{GCS}} = \eta\ M_{\mathrm{h}},
\end{equation}
with $\eta = 3.5\times10^{-5}$ \citep{Harris_2015}.

Individual cluster masses are sampled from the same log-normal mass distribution adopted for the statistical galaxy samples (see Subsection~\ref{subsec:CatalogsData}), with the turnover mass set according to host galaxy stellar mass: $M_{\rm TO} = 2 \times 10^5\ {\rm M_\odot}$ for $M_* > 10^9\ {\rm M_\odot}$ and $M_{\rm TO} = 1.6 \times 10^5\ {\rm M_\odot}$ for lower-mass dwarfs, and dispersion $\sigma_{\log M} = 0.5$ dex. Clusters are drawn iteratively until the total cluster system mass matches the target value $M_{\rm GCS,0}$ for the host galaxy. These synthetic GC populations provide present-day mass distributions for the simulated dwarf galaxies, allowing the same reconstruction procedure to be applied consistently to both the observational and simulated samples.

\begin{figure}
    \includegraphics[width=\columnwidth]{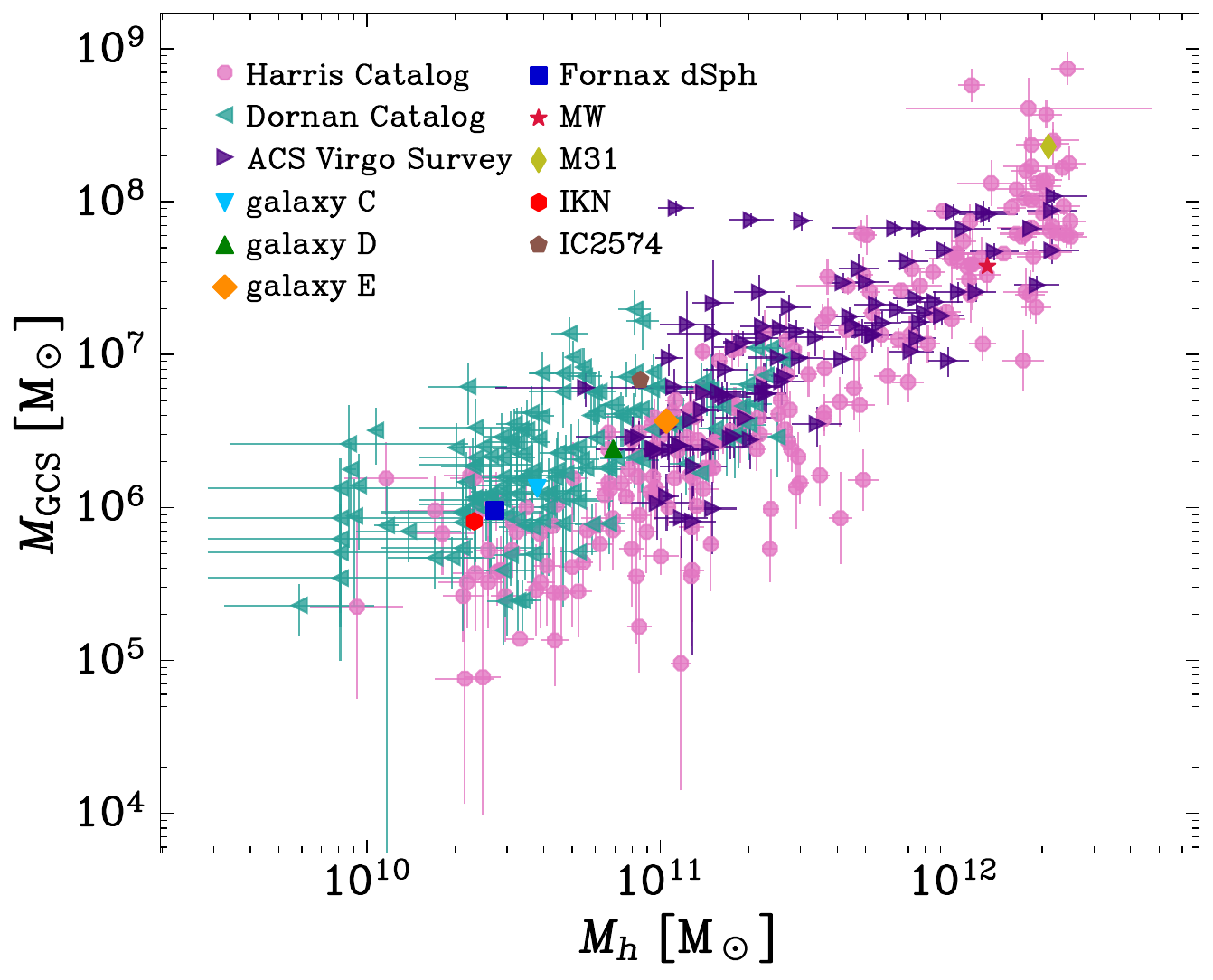}
	\centering
    \caption{Complete galaxy sample used in this work. Galaxies with directly measured GC masses, ACSVCS systems with photometrically derived GC masses, statistical catalog galaxies from \citet{Harris2013}, \citet{Dornan_2026}, and simulated dwarf galaxies are shown together in the $M_{\rm GCS}$--$M_h$ plane.}
    \label{fig:FullSample}
\end{figure}
Together, these datasets provide 451 present-day GC mass distributions across a wide range of galaxy masses and environments. These distributions form the basis for the dynamical reconstruction of the initial cluster mass function described in the following section. Figure~\ref{fig:FullSample} summarizes the complete galaxy sample used in this work, combining the systems with individual GC masses, the ACSVCS galaxies with photometrically derived GC masses, the statistical galaxy catalogs, and the simulated dwarf galaxies.

\section{Methodology}
\label{sec:Methodology}
The GCIMF at high redshift cannot be observed directly and must therefore be inferred from present-day GCSs observations and cosmological simulations. In this work, we reconstruct the high-mass end of the GCIMF of galaxies, spanning from dwarfs to Milky Way–like hosts, by linking present-day cluster masses to their corresponding progenitor values through inversion of a host-galaxy-dependent mass-loss model calibrated on direct $N$-body simulations (Paper I).

A key aspect of our approach is that no functional form for the initial cluster mass distribution is assumed. Instead, each present-day cluster mass is treated individually and inverted to recover the corresponding progenitor mass. The reconstructed GCIMF therefore emerges directly from the data rather than being imposed a priori.

The reconstruction procedure is identical for all datasets. Each present-day cluster mass, either directly observed or stochastically generated, is inverted individually using the same dynamical evolution model. The only distinction between samples lies in how the present-day mass distribution was obtained, as described in Section~\ref{sec:Data}.

\subsection{Host-Galaxy Dependent Mass-loss Model}
The long-term dynamical mass evolution of GCs is modeled using the host-galaxy-dependent calibration introduced and tested in Paper I. In that work, we performed a suite of direct $N$-body simulations of star clusters evolving in realistic, time-dependent tidal fields representative of dwarf galaxies. The simulations followed clusters with a range of initial masses and orbital configurations over a Hubble time, allowing us to quantify the average mass-loss rate as a function of cluster mass and host galaxy mass. 

The galactic tidal fields are described by time-dependent tidal tensors extracted from the cosmological simulations. For each host galaxy, we use at most two tidal histories, corresponding to clusters on different orbits within the same galaxy. As shown in Paper I, tidal tensors associated with clusters within the same galaxy can exhibit substantially different amplitudes and temporal variations, reflecting differences in their orbital histories and interactions with the galactic environment. Thus, the mass-loss history is not expected to be identical for all clusters within a given galaxy or halo mass. The tidal histories adopted here are instead intended to provide representative examples of the range of tidal environments experienced by clusters at a given host halo mass.

Table~\ref{tab:GCInitCond} summarizes the initial conditions adopted for all simulation sets used in the calibration. All models were initialized as Plummer spheres with the same initial half-mass density of $\rho_{h,0} = 10^3$ M$_\odot/$pc$^3$, while the initial particle number, and consequently the cluster mass and half-mass radius, varied among the different simulation sets. Stellar masses were sampled from a \citet{2001MNRAS.322..231K} initial mass function between $0.08$ and $100\,M_\odot$. The simulations were performed with {\nbodysix}~\citep{Wang2015}, which combines direct $N$-body integration with regularization algorithms for accurately treating close encounters and compact-object dynamics. Stellar evolution was modeled using the single and binary stellar evolution SSE/BSE prescriptions, including fallback-dependent natal kicks for neutron stars and black holes (BHs) following \citet{Banerjee2020}.
Retained stellar-mass BHs have been shown to enhance the long-term evaporation of GCs by injecting energy into the stellar component through dynamical interactions, thereby accelerating two-body relaxation and stellar escape \citep{Banerjee2011, Gieles2011, Giersz2019, Wang2020, Kremer2020, Gieles2023}. Because BH formation, retention, and dynamical evolution are treated self-consistently in our simulations, the calibrated mass-loss relation naturally incorporates the effects of BH-driven heating together with stellar evolution, two-body relaxation, and the external tidal field.
\begin{deluxetable}{lcccc}
\tablewidth{0pt}
\tablecaption{Initial conditions for the globular cluster $N$-body simulations.\label{tab:GCInitCond}}
\tablehead{
\colhead{Parameter} & \colhead{Set 1} & \colhead{Set 2} & \colhead{Set 3} & \colhead{Set 4}
}
\startdata
Number of particles, $N_0$ &
32,000 & 64,000 & 96,000 & 200,000 \\
Cluster mass, $M_0$ [$M_\odot$] &
20,480 & 40,960 & 61,440 & 128,000 \\
Half-mass radius, $r_{h0}$ [pc] &
1.34 & 1.69 & 1.94 & 2.48 \\
Stellar IMF [$M_\odot$] & \multicolumn{4}{c}{\citet{2001MNRAS.322..231K}, $0.08-100$} \\
Half-mass density, $\rho_{h,0}$ ($M_\odot\mathrm{pc}^{-3}$) & \multicolumn{4}{c}{$10^3$} \\
Density profile & \multicolumn{4}{c}{Plummer} \\
\enddata
\tablecomments{A total of 32 $N$-body simulations were performed, consisting of four sets of eight simulations each.}
\end{deluxetable}

\begin{figure}
    \includegraphics[width=\columnwidth]{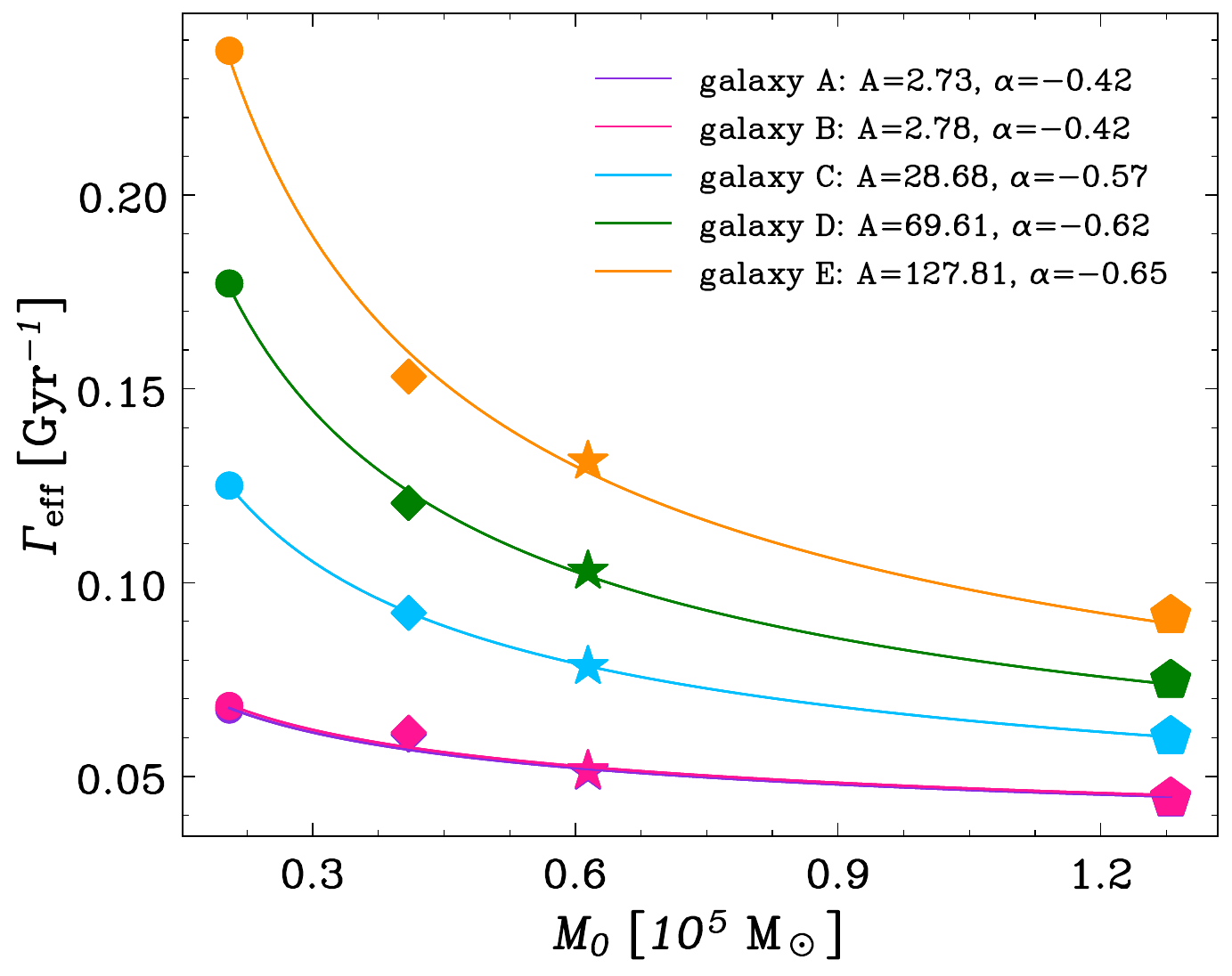}
	\centering
    \caption{Calibration of the average fractional mass-loss rate as a function of initial cluster mass, $M_0$, for the host dwarf galaxies from Paper I. Symbols indicate results from direct $N$-body simulations across multiple sets, including set~1 (circles), set~2 (diamonds), and set~3 (stars) from Paper I, as well as the latest higher-resolution set~4 (pentagons) with $N_0 = 2\times10^5$ particles. Solid lines show the power-law fits defined by Equation~\ref{eq:mdot}. These host-galaxy-dependent relations are used to model the long-term evolution of the clusters and reconstruct their progenitor masses from present-day observations.}
    \label{fig:fractional_mass_loss_calibration}
\end{figure}
To characterize the mass evolution of an individual cluster, we define its remaining mass fraction as
\begin{equation}
    \mu(t)\equiv\frac{M(t)}{M_0},
    \label{eq:remaining_mass_fraction}
\end{equation}
where $M_0$ is the initial cluster mass and $M(t)$ is its mass at time $t$. The fraction of the initial mass lost by time $t$ is $1-\mu(t)$. We characterize this evolution through the average fractional mass-loss rate
\begin{equation}
    \Gamma_{\rm eff} \equiv \frac{1-\mu(t)}{t} = \frac{1}{t}\frac{M_0-M(t)}{M_0},
\label{eq:fractional_mdot}
\end{equation}
which has units of inverse time. We define the dissolution time as $t_{\rm diss}$. 
For clusters that dissolve within $12$ Gyr ($t_{\rm diss}<12$ Gyr), Equation~\ref{eq:fractional_mdot} is evaluated at $t=t_{\rm diss}$. Clusters are considered dissolved when they retain 7 per cent of their initial mass, such that $\mu(t_{\rm diss})=M(t_{\rm diss})/M_0=0.07$. For clusters with longer lifetimes ($t_{\rm diss}>12$ Gyr), the rate is instead evaluated at $t=12$ Gyr, corresponding to a representative age of present-day GCs.

For each dwarf galaxy, the fractional mass-loss rates measured in the simulations are fitted as a function of initial cluster mass. We write
\begin{equation}
    \Gamma_{\rm eff}(M_0) = A\ {\rm Gyr}^{-1}\left(\frac{M_0}{M_\odot}\right)^\alpha,
\label{eq:mdot}
\end{equation}
where $A$ is the dimensionless numerical normalization of the
fractional mass-loss rate and $\alpha$ is the dimensionless exponent describing its dependence on initial cluster mass. Together, $(A,\alpha)$ provide an empirical description of the mass-loss trends exhibited by the simulations (Figure~\ref{fig:fractional_mass_loss_calibration}) and encode their dependence on both the initial cluster mass and the strength of the host-galaxy tidal field.

The corresponding dimensional effective mass-loss rate is
\begin{equation}
    \left\langle|\dot M|\right\rangle_{\rm eff} = M_0\Gamma_{\rm eff}(M_0) = A M_\odot\ {\rm Gyr}^{-1} \left(\frac{M_0}{M_\odot}\right)^{\alpha+1}.
\label{eq:dimensional_mass_loss_rate}
\end{equation}
Thus, $\Gamma_{\rm eff}$ describes the effective mass-loss rate normalized by the initial cluster mass, while $\langle|\dot M|\rangle_{\rm eff}$ gives the corresponding mass lost per unit time, in $M_\odot\,{\rm Gyr}^{-1}$.

In this work we adopt the calibrated model from Equation~\ref{eq:mdot} to describe the long-term mass evolution of GC systems. As described in Table~\ref{tab:GCInitCond}, in addition to the simulations presented in Paper I, we include an extended set of simulations (hereafter set 4) with increased initial particle numbers of $N_0 = 2\times10^5$ particles per star cluster extending the calibration to $M_0=1.28\times10^5\ M_\odot$, as shown in Figure~\ref{fig:fractional_mass_loss_calibration}. These higher-resolution simulations improve the sampling of the mass-loss relation and allow for a more robust determination of the parameters $(A,\alpha)$ across different galactic environments.

Approximating the effective fractional mass-loss rate of each cluster as constant over the adopted evolution interval, its remaining mass fraction at time $t$ is
\begin{equation}
    \mu(t) = 1-\Gamma_{\rm eff}(M_0)t = 1- A\left(\frac{M_0}{M_\odot}\right)^\alpha \left(\frac{t}{\rm Gyr}\right).
\label{eq:forward_mass_fraction}
\end{equation}
The corresponding cluster mass is
\begin{equation}
    M(t) = M_0\mu(t) = M_0 \left[1- A\left(\frac{M_0}{M_\odot}\right)^\alpha \left(\frac{t}{\rm Gyr}\right) \right].
\label{eq:forward_cluster_mass}
\end{equation}
Throughout this work, we adopt $t=12$ Gyr when relating progenitor masses to the masses of surviving present-day GCs.

For each observed present-day cluster mass $M_{\rm obs}$, we reconstruct the progenitor mass by numerically solving
\begin{equation}
    M_{\rm obs} = M_0 \left[1- A\left(\frac{M_0}{M_\odot}\right)^\alpha \left(\frac{t}{\rm Gyr}\right) \right]
\label{eq:inverse_progenitor_mass}
\end{equation}
for $M_0$. The solution for $M_0$ is obtained using a Brent root-finding algorithm over the relevant initial-mass range.

This formulation assumes that the average fractional mass-loss rate remains approximately constant over the lifetime of the cluster, such that cumulative mass loss scales linearly with time. The calibration parameters $(A,\alpha)$ are obtained by fitting the mass evolution measured from the $N$-body simulations. Since these simulations include stellar evolution, two-body relaxation, and tidal perturbations, the resulting parametrization captures the combined effect of these processes on cluster mass evolution. 

This host-galaxy-dependent calibration provides the key link between present-day cluster masses and their progenitor values, enabling the empirical reconstruction of GCIMFs. Appendix~\ref{app:GG23_comparison} compares the resulting effective mass-loss rates with the cluster-evolution prescription of \citet{Gieles2023}, placing our results in the context of previous work.

\subsection{Halo Mass Assignment and Mass-Loss Parameter Scaling}
For catalog galaxies with stellar mass estimates, halo masses are assigned using the UniverseMachine DR1 posterior mean stellar–halo mass relation $\langle \log M_{\rm peak} | \log M_* \rangle$ at $z \approx 0$ \citep{Behroozi_2019}. Satellite galaxies are assigned their peak halo mass. Uncertainties include intrinsic scatter in the stellar–halo mass relation combined in quadrature with stellar-mass measurement uncertainties.

From our $N$-body simulations, we find that the mass-loss parameters $(A,\alpha)$ follow power-law relations as a function of halo mass. We therefore propose scaling relations of the form
\begin{equation}
\log A = a_1 \log \left(\frac{M_{\rm h}}{M_\odot}\right) + b_1,
\end{equation}
\begin{equation}
\alpha = a_2 \log \left(\frac{M_{\rm h}}{M_\odot}\right) + b_2,
\end{equation}
where $(a_1=1.129, b_1=-8.398)$ and $(a_2=-0.157, b_2=1.091)$ are the best-fit slope and intercept parameters for the $\log A$ - $\log M_{\rm h}$ and $\alpha$ - $\log M_{\rm h}$ relations, respectively. These relations allow us to predict environmentally dependent mass-loss rates for each galaxy in the sample.

\subsection{Reconstruction of the GCIMF}
Each present-day cluster mass, either directly observed or statistically generated, is inverted using the appropriate $(A,\alpha)$ parameters to recover its progenitor mass. The reconstructed GCIMF is then built in logarithmic mass bins. We characterize the high-mass end of the distribution by fitting a power law of the form
\begin{equation}
\frac{dN}{d\log M} \propto M^{-\beta}
\label{eq:GCIMF_slope}
\end{equation}
where $\beta$ is the high-mass slope of the reconstructed GCIMF. The fit is performed in log-log space above the turnover region. Uncertainties on $\beta$ are estimated from the covariance matrix of the linear regression.

Because the reconstruction begins from present-day GCs, it necessarily includes only systems that survive to $z=0$ and excludes clusters that were fully disrupted during their dynamical evolution. The recovered distribution therefore represents the progenitor masses of surviving clusters rather than the full initial cluster population. As a consequence, the reconstructed GCIMF becomes incomplete toward low masses, where dynamical disruption is most efficient. Our method therefore primarily constrains the high-mass range of the GCIMF, corresponding to the progenitors of clusters that survived to the present day. The low-mass portion of the initial cluster population, which is expected to be dominated by clusters that were disrupted over cosmic time, cannot be recovered with this approach. Despite this limitation, the survivor-conditioned reconstruction provides direct empirical constraints on the high-mass regime of the GCIMF and on the environmental dependence of long-term dynamical evolution across diverse galactic environments.

This approach allows us to test whether the observed present-day GCMFs, characterized by their near-universal turnover mass, can arise from progenitor mass functions consistent with canonical power-law slopes, or whether the reconstructed GCIMFs exhibit systematic variations with galaxy mass. In this sense, the reconstructed distributions should be interpreted as empirical constraints on the high-mass end of the GCIMF that remains accessible through surviving clusters.

\section{Recovered GCIMF\lowercase{s}}
\label{sec:Results}
In this section we apply the reconstruction procedure described in Section~\ref{sec:Methodology} to all GC systems in our sample. This includes galaxies with individually measured GC masses, large statistical galaxy samples, and synthetic GC populations associated with simulated dwarf galaxies. For each system we reconstruct the GCIMF by inverting the mass loss evolution of present-day clusters in the system.

\begin{figure}
    \includegraphics[width=\columnwidth]{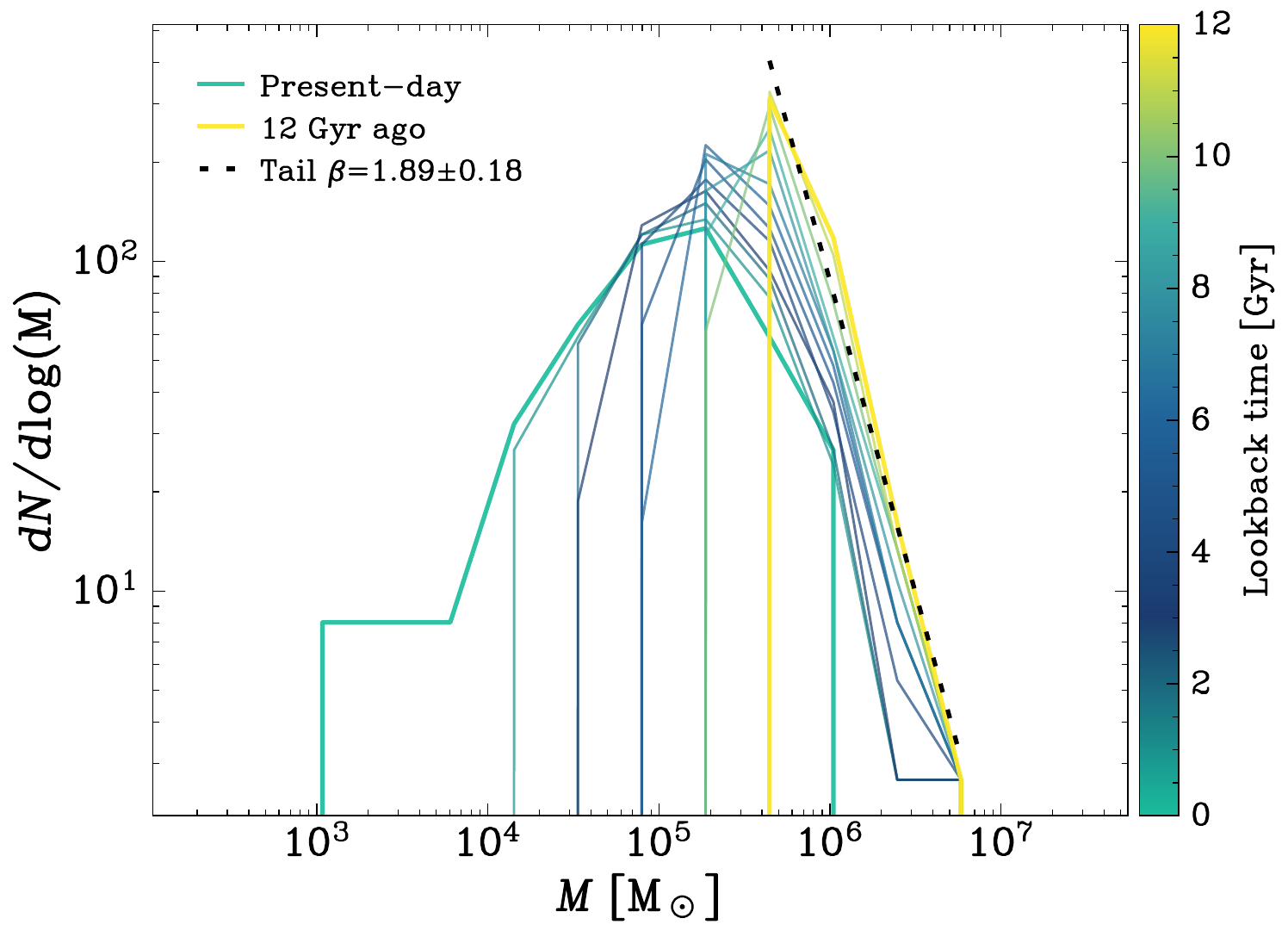}
    \caption{Reconstructed GCIMF (12 Gyr ago) for the Milky Way (yellow histogram) compared to the present-day GCMF (teal histogram) of 167 GCs. The dashed line shows the best-fit high-mass power-law slope, yielding $\beta = 1.89 \pm 0.18$ in $dN/d\log M$.}
    \label{fig:MW_GCIMF_reconstructed}
\end{figure}

We begin with the subsample of galaxies for which dynamical masses are available for individual GCs. These systems include the MW, M31, and several nearby dwarf galaxies. Because individual cluster masses are known, these galaxies provide the most direct observational test of our reconstruction procedure.
Figure~\ref{fig:MW_GCIMF_reconstructed} shows the reconstruction procedure for the Milky Way as a representative example of the galaxies with individually measured GC masses described in Subsection~\ref{subsec:IndividualGCMasses}. Starting from the present-day GCMF, we invert the mass-loss evolution of the GC system to recover the inferred progenitor mass distribution $12\ \rm Gyr$ ago, following the methodology described in Section~\ref{sec:Methodology}. Reconstruction figures for the remaining systems in this subsample, including M31 and the dwarf galaxy systems, are presented in Figures~\ref{fig:M31_GCIMF_reconstructed}, \ref{fig:IC2574_GCIMF_reconstructed}, \ref{fig:NGC3073_GCIMF_reconstructed}, and \ref{fig:dw1312p4147_GCIMF_reconstructed} in  Appendix~\ref{sec:Appendix_AdditionalGCIMFs}.

Note that, since the reconstruction method consists in gradually adding the mass lost by every GC, the reconstructed distributions are systematically shifted toward higher masses relative to the present-day GCMFs. This behavior reflects the cumulative effects of dynamical evolution over a Hubble time, which preferentially disrupts low-mass clusters and reduces the masses of surviving systems. As the evolution is inverted backward in time, the mass distributions progressively deviate from the characteristic peaked present-day GCMF and approach a power-law form at the high-mass end. During this process, intermediate timesteps can exhibit the appearance of new peaks in the reconstructed distributions. These features are artificial and arise because the reconstruction only traces surviving clusters, while systems that were fully disrupted over cosmic time are absent from the inversion. Consequently, peaks appearing at intermediate stages of the reconstructed GCIMFs should not be interpreted as physical features of the original cluster population.

\subsection{High-Mass regime slopes of the reconstructed GCIMF}
Since the main goal of this work is to characterize the GCIMF (or at least its high-mass end) for a large number of galaxies, we quantify the shape of the reconstructed GCIMF distributions using the power-law slope $\beta$ defined in Equation~(\ref{eq:GCIMF_slope}). The fit is restricted to masses above the turnover region, where the reconstructed distributions remain complete.

\begin{figure}
    \centering
    \includegraphics[width=\columnwidth]{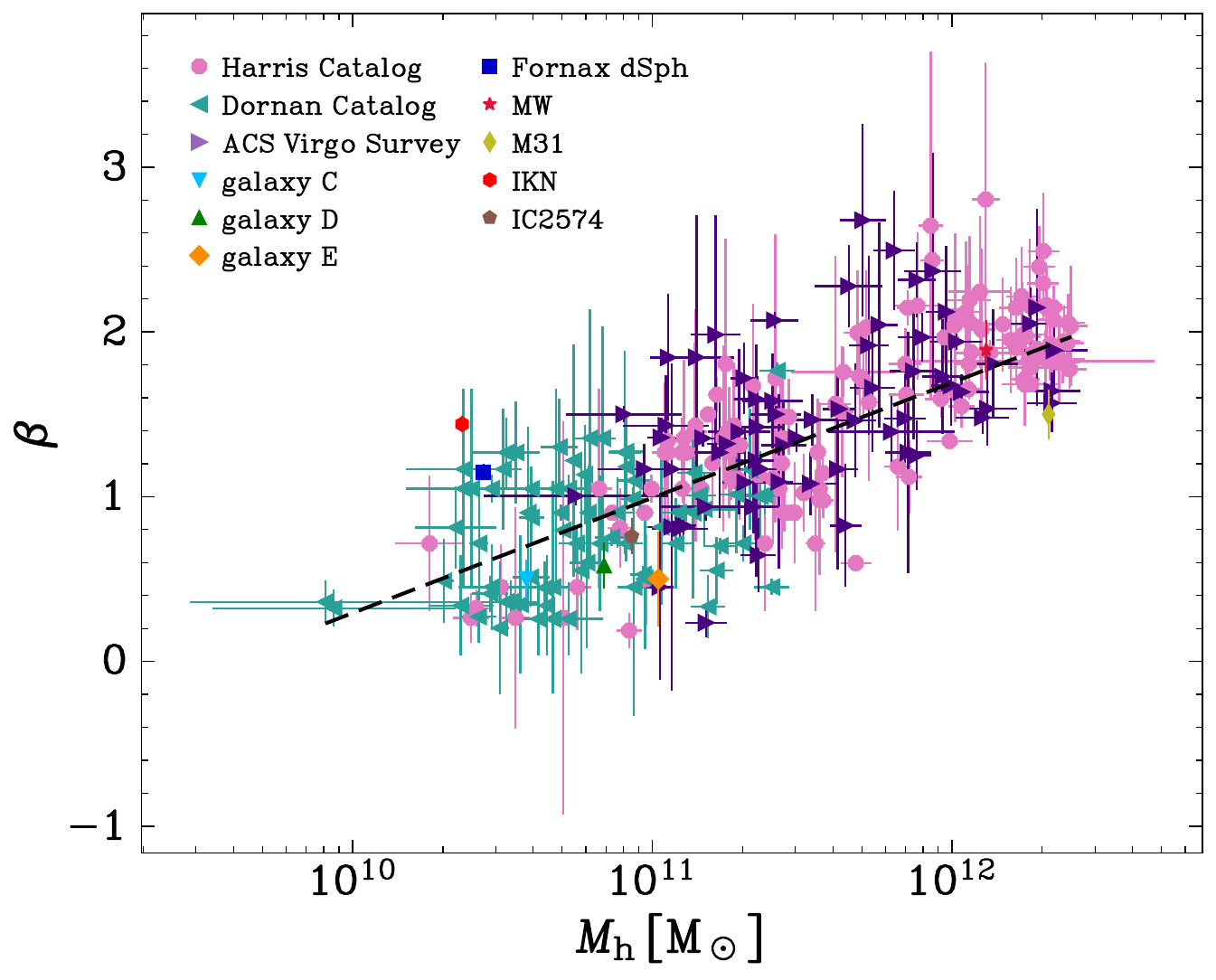}
    \caption{High-mass slope $\beta$ of the reconstructed GCIMF as a function of halo mass for the galaxy sample analyzed in this work. The dashed line shows the best-fitting linear relation. The best fit parameters are $\beta = (0.699 \pm 0.038) \ \log(M_{\rm h}) - (6.698 \pm 0.428)$, with a scatter of $\sigma = 0.368$.}
    \label{fig:fit_range}
\end{figure}

Figure~\ref{fig:fit_range} shows the inferred slopes as a function of halo mass for all galaxies where the reconstruction was possible. The resulting slopes, $\beta$, vary systematically across galaxies and show a positive correlation with host galaxy halo mass. The inferred high-mass slopes for a representative subset of galaxies, including systems with individual GC mass measurements as well as galaxies drawn from the catalogs, are listed in Table~\ref{tab:results_summary}, providing a compact overview of the variation in GCIMF properties across different galaxy environments.
To characterize the dependence of $\beta$ on the galaxy halo mass we fit a linear relation of the form
\begin{equation}
\beta = a\ \log\left(\frac{M_{\rm h}}{M_\odot}\right) + b,
\end{equation}
where $a$ and $b$ are the slope and intercept of the best-fitting linear relation, respectively. The fit is performed using the subset of galaxies for which a reliable high-mass GCIMF slope $\beta$ could be measured from the reconstruction. Of the 451 total galaxies in our compiled sample, 274 yielded enough statistics for the linear regression; the rest were excluded because their present-day GC populations were either too small to allow a fit (e.g., only a single histogram bin) or produced mass function histograms that, while technically fit-able, contained very few clusters per bin (e.g., three bins with a single GC each), making the resulting high-mass slope measurements unreliable. The $\beta$ fit was performed over the halo mass range $\log(M_{\rm h}/{\rm M_\odot}) = 9.9$ - $12.4$.

The regression was performed using an Orthogonal Distance Regression (ODR) method, which accounts for uncertainties in both the inferred slope $\beta$ and the host halo mass $M_{\rm h}$. Unlike ordinary least-squares fitting, ODR minimizes the orthogonal distances of the data points from the best-fit relation and therefore provides a more appropriate treatment when both variables carry measurement uncertainties. 

The resulting best-fit relation is
\begin{equation}
\beta = (0.699 \pm 0.038) \log\left(\frac{M_{\rm h}}{10^{11} \rm M_\odot}\right)+(0.993 \pm 0.025)
\end{equation}
where the halo mass is normalized to a representative value of $10^{11}\,{\rm M_\odot}$, close to the center of the fitted halo-mass range. The scatter of the data around the linear relation, $\sigma = 0.368$, reflects the galaxy to galaxy variation in $\beta$, showing that galaxies of similar halo mass can host GC systems with slightly different high-mass slopes. This confirms that the linear fit captures the average behavior and establishes that more massive galaxies tend to host reconstructed GCIMFs with steeper high-mass slopes.
\begin{deluxetable}{lccc}
\tablewidth{\textwidth}
\tablecaption{Reconstructed GCIMF properties for a representative galaxy sample. \label{tab:results_summary}}
\tablehead{
\colhead{Galaxy} & \colhead{$M_{\mathrm{h}}$} & \colhead{$N_{\mathrm{GC}}$} & \colhead{$\beta$} \\
\colhead{} & \colhead{$(M_\odot)$} & \colhead{} & \colhead{}
}
\startdata
Milky Way & $1.3 \times 10^{12}$   & 167  & $1.89 \pm 0.18$ \\
M31       & $2.1 \times 10^{12}$   & 436  & $1.50 \pm 0.15$ \\
NGC3073   & $4.34 \times 10^{11}$  & 130  & $1.76 \pm 0.15$ \\
IC2574    & $8.54 \times 10^{10}$  & 23   & $0.76 \pm 0.11$ \\
UGCA337   & $5.25 \times 10^{10}$  & 10   & $1.05 \pm 0.08$ \\
\enddata
\tablecomments{Host halo masses ($M_{\mathrm{h}}$), number of globular clusters ($N_{\mathrm{GC}}$), and inferred high-mass slope ($\beta$) of the reconstructed GCIMF.}
\end{deluxetable}

\section{Discussion and Conclusions}
\label{sec:DiscussionAndConclusions}
We have presented a method to reconstruct the GCIMF from present-day cluster populations, accounting for dynamical mass loss over cosmic time using a calibrated model based on $N$-body simulations of star clusters in realistic tidal environments. In addition to the simulations presented in Paper I, we include an additional set with increased cluster initial particle number ($N_0 = 2 \times 10^5$), improving the robustness of the mass-loss calibration. By applying this method to a large sample of galaxies spanning a wide range of masses, we find that the slope of the high-mass regime of the reconstructed GCIMF varies systematically with host galaxy mass. In particular, more massive galaxies tend to exhibit steeper reconstructed slopes, while dwarf galaxies show comparatively shallower distributions. 

Direct measurements of the GCIMF at high redshift are still challenging, as this would require observations of a large number of GC systems at $z \gtrsim 2$. While recent observations with the JWST have identified candidate proto-globular clusters at very high redshift, including systems at $z \sim 6$ - $10$ \citep[e.g.,][]{Mowla2022,Vanzella2023,Adamo2024}, these are detections of small numbers of sources per galaxy and the statistics are insufficient to probe full GC populations. Nonetheless, important efforts are being made to measure part of the GCMF with the available data at high-redshift. In a recent work, \citet{Claeyssens2026} present a sample of 222 star clusters in 78 magnified galaxies, 145 of these clusters were observed by the GLIMPSE program (JWST/NIRCam imaging of a strongly lensed galaxy field) combined with sources from the literature. Given the low numbers of high-z star clusters, the authors compile a subsample of 60 young clusters that belong to different galaxies to produce an averaged GCMF in the regime of cluster masses larger than $2 \times 10^6$ M$_\odot$. However, given the technical limitations and completeness effects involved in observe star cluster systems in individual high-z galaxies, measurements on the slope of the GCMF at high-z are still an open problem. The method we present here provides a complementary approach, combining high-resolution simulations with the observational constraints currently available, allowing us to infer properties of the progenitor GC population indirectly.

A direct comparison of our conclusions can be made with recent observational constraints on cluster formation at high redshift. \citet{Vanzella_2026} report the discovery and analysis of several massive stellar clusters in a galaxy at $z \sim 9.6$, providing a direct test of the GCIMF shape at formation. Their results suggest that a significant fraction of the  galaxy stellar mass is formed in compact clusters, and for this total cluster mass to not exceed the available mass from the host galaxy's last burst, the slope of the GCIMF must be shallower than values commonly adopted in the literature. These findings are broadly consistent with our reconstructed GCIMFs, which at the high-mass end exhibit power-law-like behavior with slopes that vary systematically across different host galaxies. The two approaches probe complementary regimes: while high-redshift observations constrain the cluster population at birth, our method infers the progenitor distribution from present-day survivors after $\sim$12 Gyr of dynamical evolution. The agreement between these independent approaches supports a picture in which GCs form with a power-law-like mass distribution at early times, which is subsequently modified by environmental effects and dynamical mass loss. At the same time, our results suggest that the degree of this modification depends on host galaxy mass.

A key aspect of the reconstruction method is that it relies exclusively on clusters that survive to the present day. GCs that were completely disrupted during their dynamical evolution are not represented in the present-day population and therefore cannot be recovered through the inversion procedure. As a consequence, the reconstructed GCIMF should be interpreted as the progenitor mass distribution of surviving clusters rather than the full initial GC population. In particular, the method constrains robustly the upper portion of the GCIMF, while the low-mass regime remains uncertain due to the preferential disruption of low-mass clusters across cosmic time.

In massive galaxies, although many low-mass clusters are disrupted, the overall formation of clusters is also more prolific, meaning that the surviving population still provides meaningful information on the GCIMF at high masses. Conversely, in low-mass dwarf galaxies, cluster disruption is relatively inefficient, as shown in Paper I, such that the present-day population retains a larger fraction of the original clusters. In these systems, the reconstructed GCIMF closely reflects the true initial distribution, and the limitations of the method are correspondingly smaller.

The observed correlation between $\beta$ and halo mass likely reflects a combination of intrinsic variations in cluster formation efficiency and differences in cluster survival fractions. At the same time, there is also a possibility that some aspects of GC evolution in a cosmological context are not yet fully captured in current models. Processes such as tidal disruption histories, galaxy assembly, and the contribution of accreted clusters may introduce additional complexity that is not fully accounted for in our method. This is particularly relevant for massive galaxies such as the Milky Way and M31, whose GC populations include both in-situ and accreted clusters. Since our mass-loss model is calibrated primarily for clusters evolving within a given host, its applicability to accreted systems is, for now, a first approximation. The reconstructed GCIMF for such galaxies should therefore be interpreted with caution, as the evolutionary histories of their cluster systems may differ from the assumptions of the model. However, these assumptions do not change the conclusion that the slopes of recovered GCIMFs are different for different galaxies.

An additional consideration in the mass-loss calibration is that all models begin with the same initial half-mass density, $\rho_{h,0}=10^3\ M_\odot\ {\rm pc}^{-3}$. Equal initial half-mass density does not, however, imply equal tidal filling factor because the tidal field strength, and hence the tidal density, varies among the five galaxies. Clusters with the same initial density can therefore experience different rates of tidal heating and stellar escape. This environmental variation is included in the present calibration through the different tidal histories. The calibration does not, however, capture how the mass evolution would change with $\rho_{h,0}$. At fixed cluster mass and tidal field, a higher initial density corresponds to a smaller half-mass radius and a more underfilling cluster. This would reduce the initial susceptibility of the cluster to tidal truncation and shocks. At the same time, the shorter half-mass relaxation time could accelerate internal evolution and the segregation and ejection of stellar-mass BHs, thereby changing the expansion and mass loss of the cluster. Quantifying the effect of increasing the initial density would require simulations spanning both initial density and tidal environment.

An additional caveat arises from the reconstruction procedure itself. Because only surviving clusters are included, the mass distributions obtained when evolving the system backward in time exhibit an artificial peak. This feature reflects the absence of disrupted clusters rather than a physical property of the GCIMF. Consequently, the low-mass and peak mass regions of the reconstructed distributions should not be overinterpreted. In contrast, the high-mass regime (masses larger than the peak), provides a more reliable probe of the GCIMF shape.

Nonetheless, the approach presented here represents a novel methodology for inferring progenitor GC mass distributions directly from present-day cluster populations, combining the available observational properties and constrains of GC systems with physically motivated dynamical models. An important advantage of this approach is that it does not assumes a functional form for the GCIMF, that although prevents us from recovering the full GCIMF, it leaves the recovered part model-independent. Future work should extend this analysis to larger galaxy samples and incorporate models that account for the disruption of clusters no longer present today, which will help constrain the full initial cluster population and the environmental dependence of GC formation and evolution.

Looking forward, an important challenge is to incorporate the contribution of fully disrupted clusters and to model GC evolution within a fully cosmological framework. While some simulations already include star by star cluster formation, they are often limited to isolated galaxies. Extending these models to cosmological scales, where both galaxy assembly and cluster evolution are treated self-consistently, remains a major goal for future work.

In summary, our main findings are as follows:
\begin{enumerate}
    \item High-mass slopes $\beta$ of the reconstructed GCIMF correlate with host galaxy halo mass, with more massive galaxies exhibiting steeper slopes, while dwarf galaxies have shallower distributions.
    \item The reconstruction method primarily constrains the high-mass portion of the GCIMF, while the low-mass clusters that were disrupted are not recovered, meaning the reconstruction recovers the true initial population of the high-mass portion.
    \item In dwarf galaxies, where cluster disruption is inefficient, the reconstructed GCIMF closely approximates the true initial distribution, demonstrating the robustness of the method in low-mass environments.
    \item This study introduces a novel approach to link observed present-day GCs with their progenitor populations, providing a bridge between local GC systems and their high-redshift origins.
\end{enumerate}

These results suggest that the commonly assumed universal GCIMF may not hold across all galaxy types and masses. Instead, the initial cluster mass distribution and/or the efficiency of cluster survival may depend systematically on host galaxy properties. This has important implications for models of GC formation in cosmological simulations and for the interpretation of high-redshift proto-globular cluster candidates observed with the JWST. Future work incorporating cluster disruption histories and improved environmental modeling will be essential to fully constrain the GCIMF.

\begin{acknowledgments}
We thank the anonymous referee for a careful review and insightful suggestions that contributed to improve this work. We thank Oleg Gnedin and Florent Renaud for helpful discussions. EMH acknowledges support by the Tsinghua Shui Mu Scholarship and by the Beijing Natural Science Foundation with grant No. IS25036. HL is supported by the National Key R\&D Program of China No. 2023YFB3002502, the National Natural Science Foundation of China under No. 12373006 and 12533004, and the China Manned Space Program with grant No. CMS-CSST-2025-A10. LAMM acknowledges support from DGAPA-PAPIIT IN108924 grant. The authors acknowledge UNAM for providing HPC resources through grant number 6: Proyecto de investigación en la Nube UNAM-AWS.
\end{acknowledgments}

%

\software{ \texttt{NumPy} \citep{van-der-Walt:2011aa}, \texttt{Matplotlib} \citep{Hunter:2007aa},
\texttt{pandas} \citep{mckinney-proc-scipy-2010, reback2020pandas},
\texttt{SciPy} \citep{Virtanen:2020aa}.
}


\appendix
\restartappendixnumbering
\section{Additional Reconstructed GCIMFs}
\label{sec:Appendix_AdditionalGCIMFs}
This appendix presents the reconstructed GCIMFs for the remaining galaxies with individually measured GC masses discussed in Subsection~\ref{subsec:IndividualGCMasses}. These figures complement the Milky Way example shown in Section~\ref{sec:Results} and illustrate the reconstruction procedure across different galaxy environments and mass scales, including M31 and nearby dwarf galaxy systems.
\begin{figure}
    \centering
    \includegraphics[width=\columnwidth]{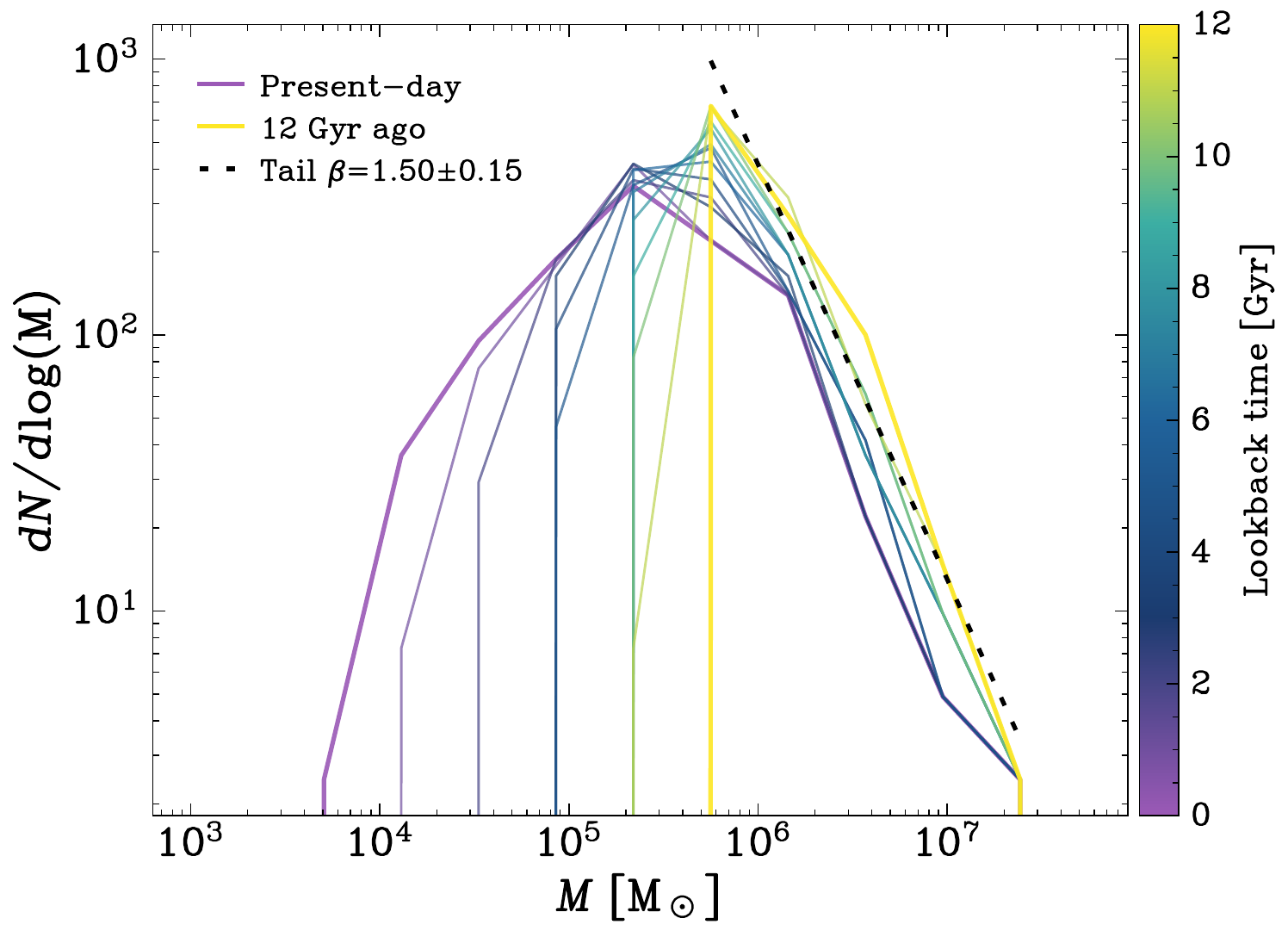}
    \caption{Reconstructed GCIMF (12 Gyr ago) for M31 (yellow histogram) compared to the present-day GCMF (purple histogram) of 436 GCs. The dashed line shows the best-fit high-mass power-law slope, yielding $\beta = 1.50 \pm 0.15$ in $dN/d\log M$.}
    \label{fig:M31_GCIMF_reconstructed}
\end{figure}
\begin{figure}
    \centering
    \includegraphics[width=\columnwidth]{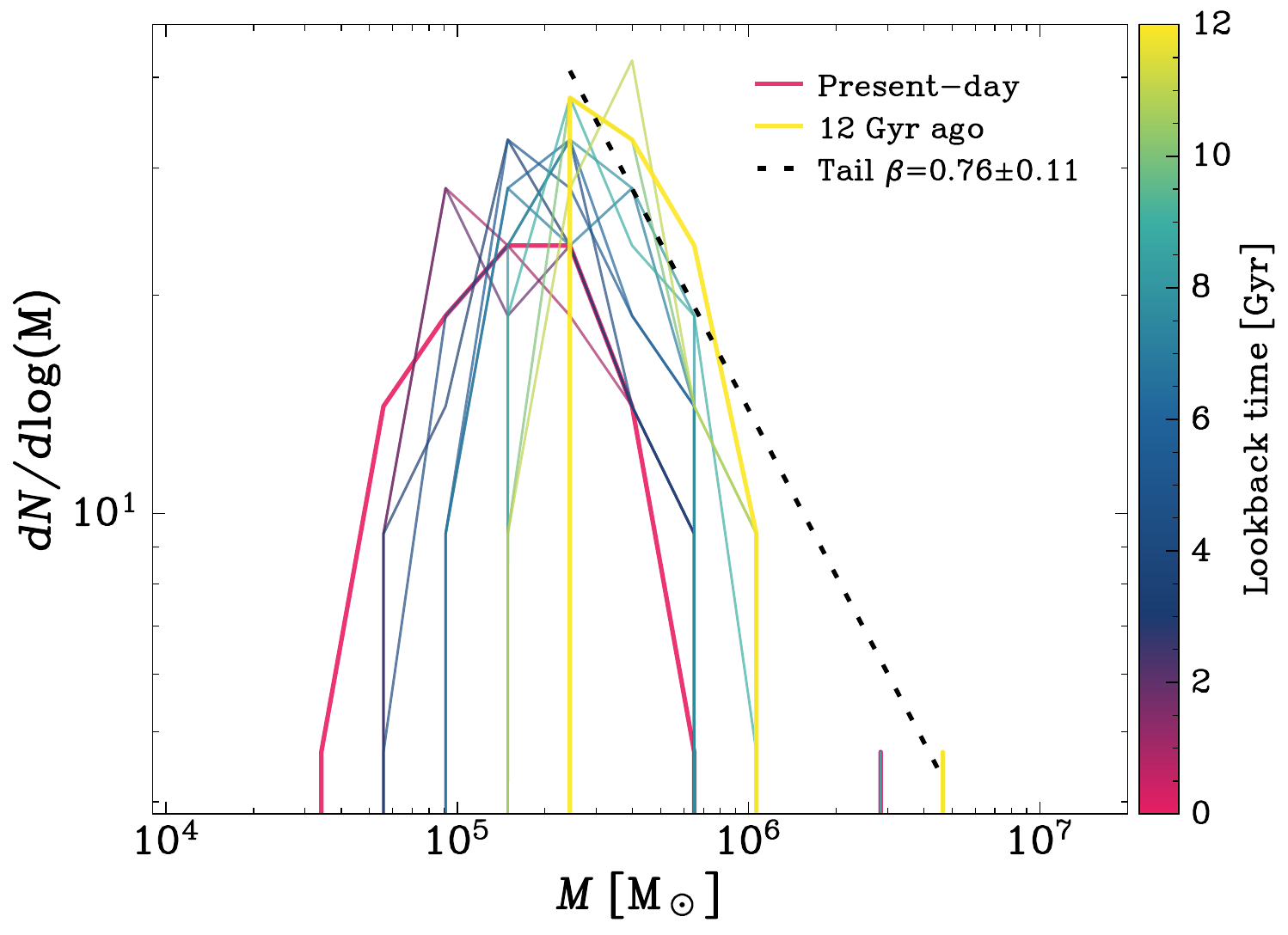}
    \caption{Reconstructed GCIMF (12 Gyr ago) for IC2574 (yellow histogram) compared to the present-day GCMF (pink histogram) of 23 GCs. The dashed line shows the best-fit high-mass power-law slope, yielding $\beta = 0.76 \pm 0.11$ in $dN/d\log M$.}
    \label{fig:IC2574_GCIMF_reconstructed}
\end{figure}
\begin{figure}
    \centering
    \includegraphics[width=\columnwidth]{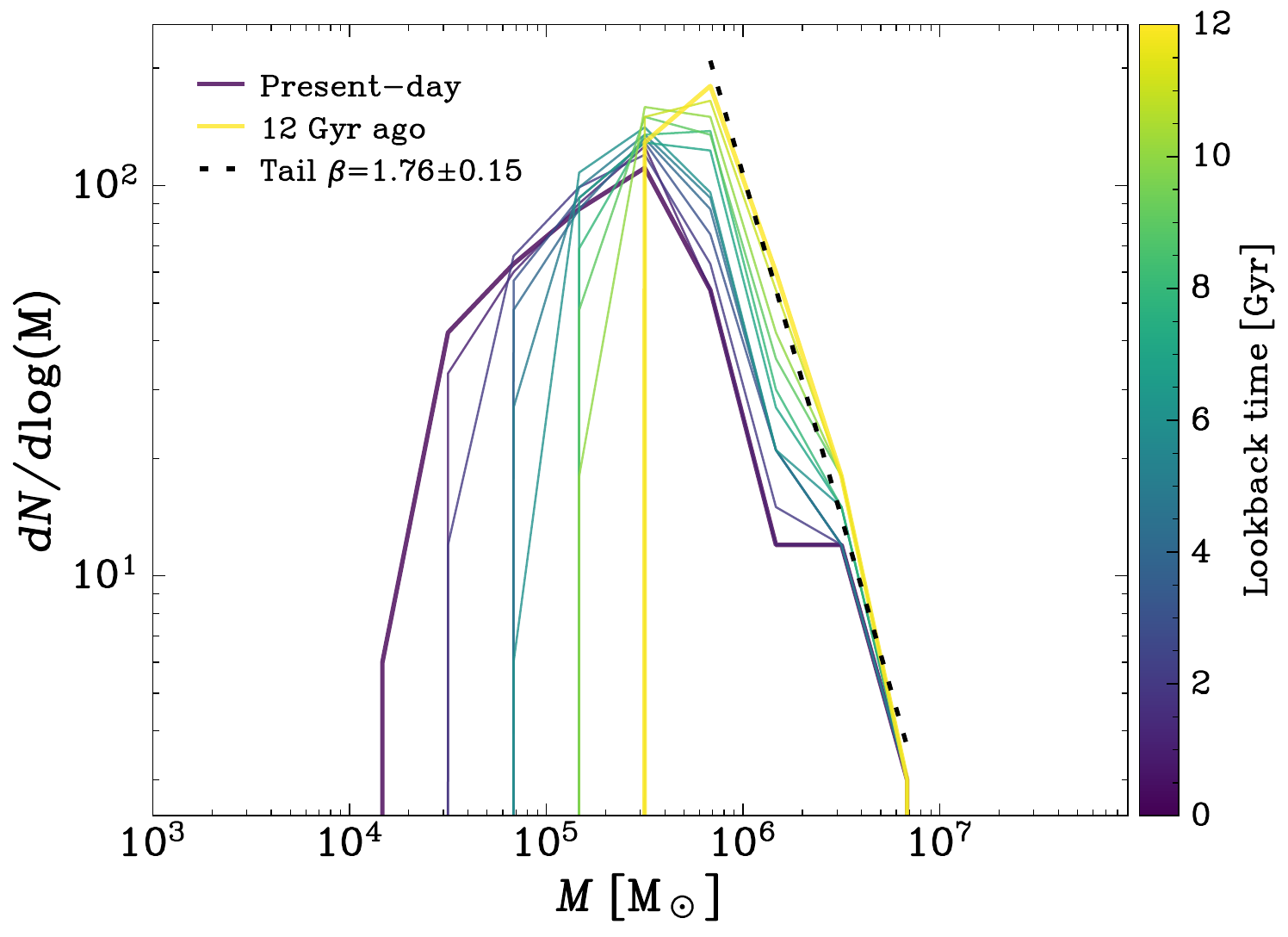}
    \caption{Reconstructed GCIMF (12 Gyr ago) for NGC3073 (yellow histogram) compared to the present-day GCMF (purple histogram). This system is taken from the GC system compilation of \citet{Harris2013} and contains 130 GCs. The dashed line shows the best-fit high-mass power-law slope, yielding $\beta = 1.76 \pm 0.15$ in $dN/d\log M$.}
    \label{fig:NGC3073_GCIMF_reconstructed}
\end{figure}
\begin{figure}
    \centering
    \includegraphics[width=\columnwidth]{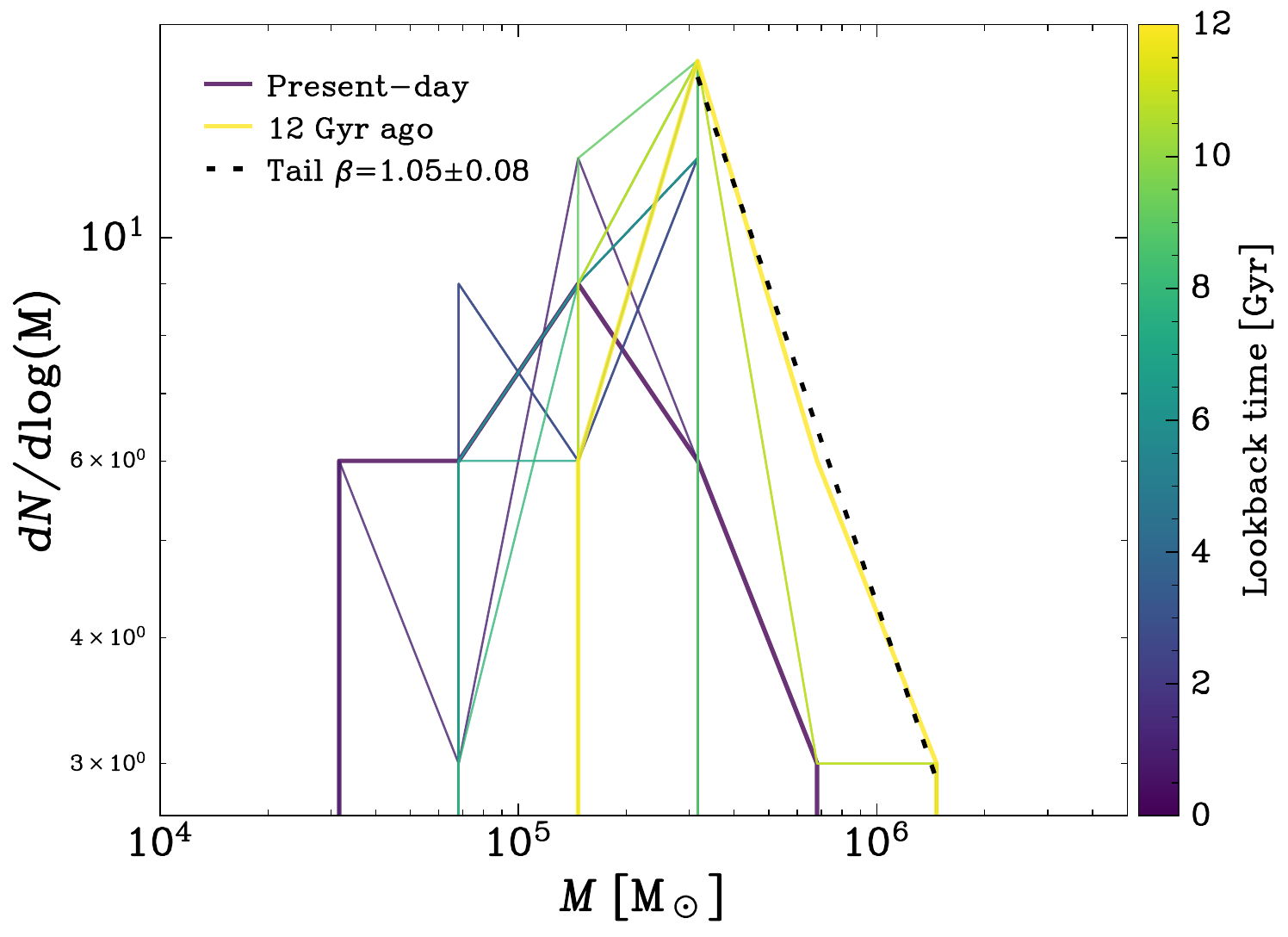}
    \caption{Reconstructed GCIMF (12 Gyr ago) for the dwarf galaxy UGCA337 (yellow histogram) compared to the present-day GCMF (purple histogram). This system is taken from the GC system compilation of \citet{Dornan_2026} and contains 10 GCs. The dashed line shows the best-fit high-mass power-law slope, yielding $\beta = 1.05 \pm 0.08$ in $dN/d\log M$.}
    \label{fig:dw1312p4147_GCIMF_reconstructed}
\end{figure}

\section{Effective mass-loss rates as a function of tidal strength}
\label{app:GG23_comparison}

In this appendix, we compare the effective mass-loss rates used in our GCIMF reconstruction with the cluster-evolution prescription of \citet[][hereafter \citetalias{Gieles2023}]{Gieles2023}, which is expressed in terms of the galactic tidal field. The comparison provides an external consistency check and does not alter our results for the reconstructed GCIMFs.

\subsection{Representative tidal strength and initial tidal filling}

In order to present our mass-loss rates in terms of the tidal strength, at every output of a tidal history we symmetrize the tidal tensor, label its eigenvalues using the convention $\lambda_1\geq\lambda_2\geq\lambda_3$, and adopt the \citetalias{Gieles2023} definition
\begin{equation}
    \Omega_{\rm tid}(t)=\sqrt{\lambda_1(t)-\lambda_3(t)}.
\label{eq:GG23_omega}
\end{equation}
We characterize each time-dependent tidal history by its median value,
\begin{equation}
    \Omega_{\rm med}\equiv {\rm median}_t\left[\Omega_{\rm tid}(t)\right].
\label{eq:median_omega}
\end{equation}
For host galaxies with two orbital histories, we average their two independently measured medians. The resulting values for host galaxies A - E are $12.73$, $11.87$, $21.31$, $37.45$, and $41.59\ {\rm Gyr}^{-1}$, respectively.

The \citetalias{Gieles2023} mass-loss prescription also depends on the initial cluster density relative to the tidal density. They characterize this dependence using the dimensionless parameter
\begin{equation}
    \mathcal R_0 \equiv \frac{\rho_{h,0}}{\rho_{h,f}},
    \label{eq:GG23_R0_definition}
\end{equation}
where $\rho_{h,0}$ is the initial half-mass density and $\rho_{h,f}$ is the half-mass density of a nominally Roche-filling cluster.

Under the assumption that a galaxy has a flat rotation curve, the effective Jacobi density associated with the representative tidal frequency is
\begin{equation}
    \rho_{J,\rm eff} = \frac{3}{2\pi G}\Omega_{\rm med}^2.
    \label{eq:GG23_Jacobi_density}
\end{equation}

Then, \citetalias{Gieles2023} define a nominally Roche-filling cluster as one with $r_h/r_J=0.145$. The corresponding half-mass density is
\begin{equation}
    \rho_{h,f} = \frac{1}{2}(0.145)^{-3}\rho_{J,\rm eff}.
    \label{eq:GG23_filling_density}
\end{equation}

Using the representative tidal frequencies computed above, and since all our $N$-body models begin with $\rho_{h,0}=10^3\ M_\odot\ {\rm pc}^{-3}$, we obtain $\mathcal R_0\simeq355$, $408$, $126$, $41$, and $33$ for galaxies A - E, respectively. Thus, although all clusters begin with the same initial half-mass density, their effective degrees of tidal filling differ because the tidal-field strength varies among the host galaxies.

The \citetalias{Gieles2023} simulations include models with $\mathcal R_0=2.9, 9.5, 29, 95$, and compared to these values our models for galaxies A and B are substantially more underfilling. On the other hand, models in galaxies D and E are closest to their $\mathcal R_0\simeq29$ model. Meanwhile, models in galaxy C ($\mathcal R_0=126$) are closer to the \citetalias{Gieles2023} $\mathcal R_0=95$ model. We therefore use the $\mathcal R_0=29$ and $95$ models as reference predictions rather than as exact density-matched descriptions of all five host galaxies.

\subsection{Comparison of effective mass-loss rates}

\begin{figure}
    \centering
    \includegraphics[width=\columnwidth]
    {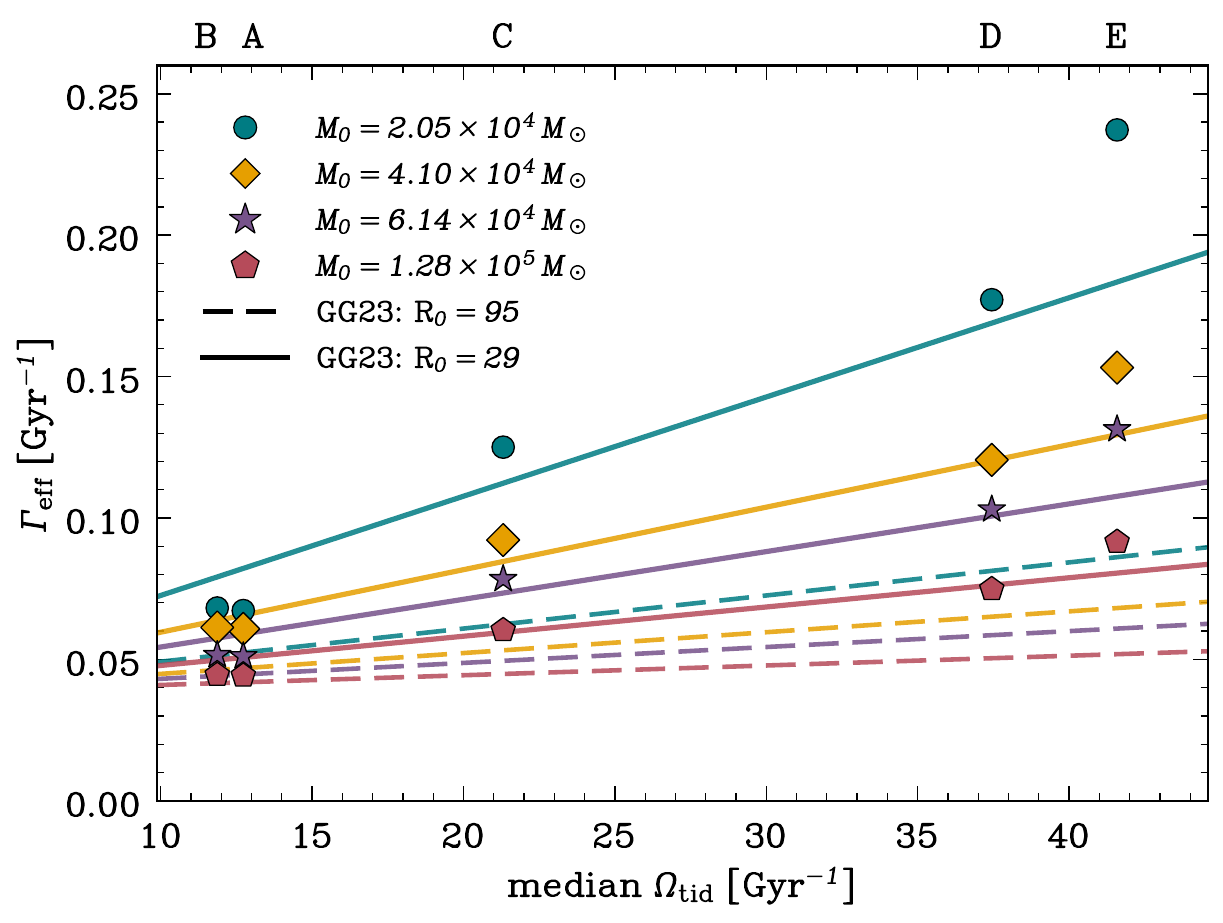}
    \caption{Total effective fractional mass-loss rates as functions of the median tidal frequency. Points show the fitted rates used in our analysis, including mass loss from stellar evolution and dynamical processes; colors and symbols distinguish the four initial cluster masses, and labels A - E identify the five host galaxies. Dashed and solid lines show the total effective \citetalias{Gieles2023} predictions for $\mathcal R_0=95$ and $\mathcal R_0=29$, respectively, using the total mass-loss rate defined in Equation~\ref{eq:GG23_total_rate}.}
    \label{fig:GG23_comparison}
\end{figure}

The effective fractional mass-loss rate used in our analysis is $\Gamma_{\rm eff}\equiv \langle|\dot M|\rangle_{\rm eff}/M_0$, where $M_0$ is the initial cluster mass. The points in Figure~\ref{fig:GG23_comparison} show the fitted
values of $\Gamma_{\rm eff}$ for the four initial cluster masses $M_0=2.05\times10^4$, $4.10\times10^4$, $6.14\times10^4$, and $1.28\times10^5\ {\rm M}_\odot$ in each of the five host galaxies. These fitted mass-loss rates include the combined effects of mass loss mechanisms present in the simulations, such as stellar evolution, internal dynamics and tidal perturbations, and are expressed as fractions of $M_0$ per Gyr.

On the other hand, \citetalias{Gieles2023} describe dynamical mass loss after the early stellar-evolution phase as
\begin{equation}
 |\dot M|_{\rm dyn}=|\dot M_{\rm ref}|
 \left(\frac{M}{M_i}\right)^{1-y}
 \left(\frac{M_i}{2\times10^5\,{\rm M_\odot}}\right)^{1-x}
 \frac{\Omega_{\rm tid}}{0.32\,{\rm Myr}^{-1}},
\label{eq:GG23_rate}
\end{equation}
where $M_i=\mu_{\rm sev}M_0$ is the cluster mass remaining after early stellar evolution. We adopt $\mu_{\rm sev}=0.55$, as in the metal-poor \citetalias{Gieles2023} models. Evaluating the \citetalias{Gieles2023} prescription using the representative tidal frequency defined in Equation~\ref{eq:median_omega}, and integrating Equation~\ref{eq:GG23_rate} from $M_i$ to zero, gives the cluster disruption time \citepalias{Gieles2023}
\begin{equation}
 t_{\rm dis}=10\ {\rm Gyr}\ 
 \frac{2/3}{y}\frac{30\ {\rm M_\odot Myr^{-1}}}{|\dot M_{\rm ref}|}
 \frac{0.32\ {\rm Myr}^{-1}}{\Omega_{\rm tid}}
 \left(\frac{M_i}{2\times10^5\ {\rm M_\odot}}\right)^x
\label{eq:GG23_tdis}
\end{equation}
Because our fitted values are constant effective mass-loss rates, we express the \citetalias{Gieles2023} dynamical evolution as the analogous lifetime-averaged rate $M_i/t_{\rm dis}$. After normalization by the original cluster mass, this gives
\begin{equation}
\Gamma_{\rm GG23,dyn} \equiv \frac{1}{M_0}\frac{M_i}{t_{\rm dis}} = \frac{\mu_{\rm sev}}{t_{\rm dis}}.
\label{eq:GG23_dyn_average}
\end{equation}
This rate has units of ${\rm Gyr}^{-1}$ when $t_{\rm dis}$ is expressed in Gyr. Following \citetalias{Gieles2023}, we adopt $x=2/3$ in Equation \ref{eq:GG23_tdis} and consider two models from their Table 1. The model with $\mathcal R_0=95$ has $y=2/3$ and $|\dot M_{\rm ref}|=30\ M_\odot{\rm Myr}^{-1}$, while the model with $\mathcal R_0=29$ has $y=4/3$ and $|\dot M_{\rm ref}|=45\ M_\odot{\rm Myr}^{-1}$.

To express these dynamical predictions using the same total effective-rate convention as our fitted values, we add the early stellar-evolution loss.
Since $M_i=0.55M_0$, this contribution corresponds to $0.45M_0$. Averaging it over the 12 Gyr interval used in our reconstruction gives
\begin{equation}
    \Gamma_{\rm sev} =\frac{1-\mu_{\rm sev}}{12\ {\rm Gyr}} =0.0375\ {\rm Gyr}^{-1}.
    \label{eq:GG23_sev_rate}
\end{equation}
The total \citetalias{Gieles2023} mass-loss rate plotted in Figure~\ref{fig:GG23_comparison} is therefore
\begin{equation}
    \Gamma_{\rm GG23,tot} =\Gamma_{\rm sev}+\Gamma_{\rm GG23,dyn}.
    \label{eq:GG23_total_rate}
\end{equation}
Both terms are normalized by $M_0$. Averaging the stellar-evolution contribution over 12 Gyr places it in the effective-rate convention used in our reconstruction; it does not imply that stellar-evolution mass loss occurs at a constant instantaneous rate.

As shown in Figure~\ref{fig:GG23_comparison}, both our fitted rates and the \citetalias{Gieles2023} predictions increase with tidal strength. Although host galaxy C has an effective density contrast that is numerically closer to $\mathcal R_0=95$, its fitted mass-loss rates lie closer to the \citetalias{Gieles2023} $\mathcal R_0=29$ predictions. On the other hand, the $\mathcal R_0$ values for galaxies D and E are closer to the $\mathcal R_0=29$ case, and Figure~\ref{fig:GG23_comparison} shows a good agreement for these galaxies and the corresponding \citetalias{Gieles2023} model. In general, the comparison illustrates that proximity in the value of $\mathcal R_0$ does not necessarily imply agreement between the mass-loss rates obtained from these different descriptions.

Note, however, that this comparison involves several simplifying choices: we represent each time-dependent tidal history by a single value of $\Omega_{\rm med}$, use this value to assign an effective $\mathcal R_0$, and express the \citetalias{Gieles2023} models as lifetime-averaged rates. The median tidal frequency does not retain information about the amplitude, duration, or frequency of individual tidal perturbations, which can also contribute to shock-driven mass loss. Moreover, the \citetalias{Gieles2023} prescription describes evolution in a steady galactic potential, whereas our fitted rates summarize evolution in time-dependent tidal fields. The \citetalias{Gieles2023} curves in Figure~\ref{fig:GG23_comparison} should therefore be interpreted as reference predictions rather than as exact descriptions of cluster evolution in each host galaxy.


\bibliography{RefLibrary}{}

@ARTICLE{Brodie_Strader2006,
       author = {{Brodie}, Jean P. and {Strader}, Jay},
        title = "{Extragalactic Globular Clusters and Galaxy Formation}",
      journal = {\araa},
         year = 2006,
        month = sep,
       volume = {44},
       number = {1},
        pages = {193-267},
          doi = {10.1146/annurev.astro.44.051905.092441},
archivePrefix = {arXiv},
       eprint = {astro-ph/0602601},
 primaryClass = {astro-ph},
       adsurl = {https://ui.adsabs.harvard.edu/abs/2006ARA&A..44..193B}
}

@ARTICLE{Jordan2007,
       author = {{Jord{\'a}n}, Andr{\'e}s and {McLaughlin}, Dean E. and {C{\^o}t{\'e}}, Patrick and {Ferrarese}, Laura and {Peng}, Eric W. and {Mei}, Simona and {Villegas}, Daniela and {Merritt}, David and {Tonry}, John L. and {West}, Michael J.},
        title = "{The ACS Virgo Cluster Survey. XII. The Luminosity Function of Globular Clusters in Early-Type Galaxies}",
      journal = {\apjs},
         year = 2007,
        month = jul,
       volume = {171},
       number = {1},
        pages = {101-145},
          doi = {10.1086/516840},
archivePrefix = {arXiv},
       eprint = {astro-ph/0702496},
 primaryClass = {astro-ph},
       adsurl = {https://ui.adsabs.harvard.edu/abs/2007ApJS..171..101J}
}

@ARTICLE{kruijssen15,
   author = {{Kruijssen}, J.~M.~D.},
    title = "{Globular clusters as the relics of regular star formation in `normal' high-redshift galaxies}",
  journal = {\mnras},
archivePrefix = "arXiv",
   eprint = {1509.02163},
     year = 2015,
    month = dec,
   volume = 454,
    pages = {1658-1686},
      doi = {10.1093/mnras/stv2026},
   adsurl = {http://adsabs.harvard.edu/abs/2015MNRAS.454.1658K}
}

@ARTICLE{Li2019,
       author = {{Li}, Hui and {Gnedin}, Oleg Y.},
        title = "{Star cluster formation in cosmological simulations - III. Dynamical and chemical evolution}",
      journal = {\mnras},
         year = 2019,
        month = jul,
       volume = {486},
       number = {3},
        pages = {4030-4043},
          doi = {10.1093/mnras/stz1114},
archivePrefix = {arXiv},
       eprint = {1810.11036},
 primaryClass = {astro-ph.GA},
       adsurl = {https://ui.adsabs.harvard.edu/abs/2019MNRAS.486.4030L}
}

@ARTICLE{Pfeffer2018,
       author = {{Pfeffer}, Joel and {Kruijssen}, J.~M. Diederik and {Crain}, Robert A. and
         {Bastian}, Nate},
        title = "{The E-MOSAICS project: simulating the formation and co-evolution of galaxies and their star cluster populations}",
      journal = {\mnras},
         year = 2018,
        month = apr,
       volume = {475},
       number = {4},
        pages = {4309-4346},
          doi = {10.1093/mnras/stx3124},
archivePrefix = {arXiv},
       eprint = {1712.00019},
 primaryClass = {astro-ph.GA},
       adsurl = {https://ui.adsabs.harvard.edu/abs/2018MNRAS.475.4309P}
}

@ARTICLE{Peng2008,
       author = {{Peng}, Eric W. and {Jord{\'a}n}, Andr{\'e}s and {C{\^o}t{\'e}}, Patrick and {Takamiya}, Marianne and {West}, Michael J. and {Blakeslee}, John P. and {Chen}, Chin-Wei and {Ferrarese}, Laura and {Mei}, Simona and {Tonry}, John L. and {West}, Andrew A.},
        title = "{The ACS Virgo Cluster Survey. XV. The Formation Efficiencies of Globular Clusters in Early-Type Galaxies: The Effects of Mass and Environment}",
      journal = {\apj},
         year = 2008,
        month = jul,
       volume = {681},
       number = {1},
        pages = {197-224},
          doi = {10.1086/587951},
archivePrefix = {arXiv},
       eprint = {0803.0330},
 primaryClass = {astro-ph},
       adsurl = {https://ui.adsabs.harvard.edu/abs/2008ApJ...681..197P}
}

@ARTICLE{Harris2013,
       author = {{Harris}, William E. and {Harris}, Gretchen L.~H. and {Alessi}, Matthew},
        title = "{A Catalog of Globular Cluster Systems: What Determines the Size of a Galaxy's Globular Cluster Population?}",
      journal = {\apj},
         year = 2013,
        month = aug,
       volume = {772},
       number = {2},
          eid = {82},
        pages = {82},
          doi = {10.1088/0004-637X/772/2/82},
archivePrefix = {arXiv},
       eprint = {1306.2247},
 primaryClass = {astro-ph.GA},
       adsurl = {https://ui.adsabs.harvard.edu/abs/2013ApJ...772...82H}
}

@article{van-der-Walt:2011aa,
	Adsurl = {https://ui.adsabs.harvard.edu/abs/2011CSE....13b..22V},
	Archiveprefix = {arXiv},
	Author = {{van der Walt}, St{\'e}fan and {Colbert}, S. Chris and {Varoquaux}, Ga{\"e}l},
	Doi = {10.1109/MCSE.2011.37},
	Eprint = {1102.1523},
	Journal = {Computing in Science and Engineering},
	Month = mar,
	Number = {2},
	Pages = {22-30},
	Primaryclass = {cs.MS},
	Title = {{The NumPy Array: A Structure for Efficient Numerical Computation}},
	Volume = {13},
	Year = 2011}

@article{Hunter:2007aa,
	Adsurl = {https://ui.adsabs.harvard.edu/abs/2007CSE.....9...90H},
	Author = {{Hunter}, John D.},
	Doi = {10.1109/MCSE.2007.55},
	Journal = {Computing in Science and Engineering},
	Month = may,
	Number = {3},
	Pages = {90-95},
	Title = {{Matplotlib: A 2D Graphics Environment}},
	Volume = {9},
	Year = 2007}

@software{reback2020pandas,
       author = {{The pandas development Team}},
        title = "{pandas-dev/pandas: Pandas}",
         year = 2025,
        month = jul,
          eid = {10.5281/zenodo.3509134},
          doi = {10.5281/zenodo.3509134},
      version = {v2.3.1},
    publisher = {Zenodo},
       adsurl = {https://ui.adsabs.harvard.edu/abs/2022zndo...3509134T}
}

@InProceedings{ mckinney-proc-scipy-2010,
  author    = { {W}es {M}c{K}inney },
  title     = { {D}ata {S}tructures for {S}tatistical {C}omputing in {P}ython },
  booktitle = { {P}roceedings of the 9th {P}ython in {S}cience {C}onference },
  pages     = { 56 - 61 },
  year      = { 2010 },
  editor    = { {S}t\'efan van der {W}alt and {J}arrod {M}illman },
  doi       = { 10.25080/Majora-92bf1922-00a }
}

@article{Virtanen:2020aa,
	Adsurl = {https://rdcu.be/b08Wh},
	Author = {Virtanen, Pauli and Gommers, Ralf and Oliphant, Travis E. and Haberland, Matt and Reddy, Tyler and Cournapeau, David and Burovski, Evgeni and Peterson, Pearu and Weckesser, Warren and Bright, Jonathan and {van der Walt}, St{\'e}fan J. and Brett, Matthew and Wilson, Joshua and Millman, K. Jarrod and Mayorov, Nikolay and Nelson, Andrew R. J. and Jones, Eric and Kern, Robert and Larson, Eric and Carey, C J and Polat, {\.I}lhan and Feng, Yu and Moore, Eric W. and {VanderPlas}, Jake and Laxalde, Denis and Perktold, Josef and Cimrman, Robert and Henriksen, Ian and Quintero, E. A. and Harris, Charles R. and Archibald, Anne M. and Ribeiro, Ant{\^o}nio H. and Pedregosa, Fabian and {van Mulbregt}, Paul and {SciPy 1.0 Contributors}},
	Doi = {10.1038/s41592-019-0686-2},
	Journal = {Nature Methods},
	Pages = {261--272},
	Title = {{{SciPy} 1.0: Fundamental Algorithms for Scientific Computing in Python}},
	Volume = {17},
	Year = {2020}}

@ARTICLE{2001MNRAS.322..231K,
       author = {{Kroupa}, Pavel},
        title = "{On the variation of the initial mass function}",
      journal = {\mnras},
         year = 2001,
        month = apr,
       volume = {322},
       number = {2},
        pages = {231-246},
          doi = {10.1046/j.1365-8711.2001.04022.x},
archivePrefix = {arXiv},
       eprint = {astro-ph/0009005},
 primaryClass = {astro-ph},
       adsurl = {https://ui.adsabs.harvard.edu/abs/2001MNRAS.322..231K}
}

@ARTICLE{Harris_2015,
       author = {{Harris}, William E. and {Harris}, Gretchen L. and {Hudson}, Michael J.},
        title = "{Dark Matter Halos in Galaxies and Globular Cluster Populations. II. Metallicity and Morphology}",
      journal = {\apj},
         year = 2015,
        month = jun,
       volume = {806},
       number = {1},
          eid = {36},
        pages = {36},
          doi = {10.1088/0004-637X/806/1/36},
archivePrefix = {arXiv},
       eprint = {1504.03199},
 primaryClass = {astro-ph.GA},
       adsurl = {https://ui.adsabs.harvard.edu/abs/2015ApJ...806...36H}
}

@ARTICLE{deBoer_2016,
       author = {{de Boer}, T.~J.~L. and {Fraser}, M.},
        title = "{Four and one more: The formation history and total mass of globular clusters in the Fornax dSph}",
      journal = {\aap},
         year = 2016,
        month = may,
       volume = {590},
          eid = {A35},
        pages = {A35},
          doi = {10.1051/0004-6361/201527580},
archivePrefix = {arXiv},
       eprint = {1510.05642},
 primaryClass = {astro-ph.GA},
       adsurl = {https://ui.adsabs.harvard.edu/abs/2016A&A...590A..35D}
}

@ARTICLE{Choksi2018,
       author = {{Choksi}, Nick and {Gnedin}, Oleg Y. and {Li}, Hui},
        title = "{Formation of globular cluster systems: from dwarf galaxies to giants}",
      journal = {\mnras},
         year = 2018,
        month = oct,
       volume = {480},
       number = {2},
        pages = {2343-2356},
          doi = {10.1093/mnras/sty1952},
archivePrefix = {arXiv},
       eprint = {1801.03515},
 primaryClass = {astro-ph.GA},
       adsurl = {https://ui.adsabs.harvard.edu/abs/2018MNRAS.480.2343C}
}

@ARTICLE{Mowla2022,
       author = {{Mowla}, Lamiya and {Iyer}, Kartheik G. and {Desprez}, Guillaume and {Estrada-Carpenter}, Vicente and {Martis}, Nicholas S. and {Noirot}, Ga{\"e}l and {Sarrouh}, Ghassan T. and {Strait}, Victoria and {Asada}, Yoshihisa and {Abraham}, Roberto G. and {Brammer}, Gabriel and {Sawicki}, Marcin and {Willott}, Chris J. and {Bradac}, Marusa and {Doyon}, Ren{\'e} and {Muzzin}, Adam and {Pacifici}, Camilla and {Ravindranath}, Swara and {Zabl}, Johannes},
        title = "{The Sparkler: Evolved High-redshift Globular Cluster Candidates Captured by JWST}",
      journal = {\apjl},
         year = 2022,
        month = oct,
       volume = {937},
       number = {2},
          eid = {L35},
        pages = {L35},
          doi = {10.3847/2041-8213/ac90ca},
archivePrefix = {arXiv},
       eprint = {2208.02233},
 primaryClass = {astro-ph.GA},
       adsurl = {https://ui.adsabs.harvard.edu/abs/2022ApJ...937L..35M}
}

@ARTICLE{Vanzella2023,
       author = {{Vanzella}, Eros and {Claeyssens}, Ad{\'e}la{\"\i}de and {Welch}, Brian and {Adamo}, Angela and {Coe}, Dan and {Diego}, Jose M. and {Mahler}, Guillaume and {Khullar}, Gourav and {Kokorev}, Vasily and {Oguri}, Masamune and {Ravindranath}, Swara and {Furtak}, Lukas J. and {Hsiao}, Tiger Yu-Yang and {Abdurro'uf} and {Mandelker}, Nir and {Brammer}, Gabriel and {Bradley}, Larry D. and {Brada{\v{c}}}, Maru{\v{s}}a and {Conselice}, Christopher J. and {Dayal}, Pratika and {Nonino}, Mario and {Andrade-Santos}, Felipe and {Windhorst}, Rogier A. and {Pirzkal}, Nor and {Sharon}, Keren and {de Mink}, S.~E. and {Fujimoto}, Seiji and {Zitrin}, Adi and {Eldridge}, Jan J. and {Norman}, Colin},
        title = "{JWST/NIRCam Probes Young Star Clusters in the Reionization Era Sunrise Arc}",
      journal = {\apj},
         year = 2023,
        month = mar,
       volume = {945},
       number = {1},
          eid = {53},
        pages = {53},
          doi = {10.3847/1538-4357/acb59a},
archivePrefix = {arXiv},
       eprint = {2211.09839},
 primaryClass = {astro-ph.GA},
       adsurl = {https://ui.adsabs.harvard.edu/abs/2023ApJ...945...53V}
}

@ARTICLE{Gieles2023,
       author = {{Gieles}, Mark and {Gnedin}, Oleg Y.},
        title = "{The mass-loss rates of star clusters with stellar-mass black holes: implications for the globular cluster mass function}",
      journal = {\mnras},
         year = 2023,
        month = jul,
       volume = {522},
       number = {4},
        pages = {5340-5357},
          doi = {10.1093/mnras/stad1287},
archivePrefix = {arXiv},
       eprint = {2303.03791},
 primaryClass = {astro-ph.GA},
       adsurl = {https://ui.adsabs.harvard.edu/abs/2023MNRAS.522.5340G}
}

@ARTICLE{Pace_2024,
       author = {{Pace}, Andrew B},
        title = "{The Local Volume Database: a library of the observed properties of nearby dwarf galaxies and star clusters}",
      journal = {The Open Journal of Astrophysics},
         year = 2025,
        month = sep,
       volume = {8},
          eid = {142},
        pages = {142},
          doi = {10.33232/001c.144859},
archivePrefix = {arXiv},
       eprint = {2411.07424},
 primaryClass = {astro-ph.GA},
       adsurl = {https://ui.adsabs.harvard.edu/abs/2025OJAp....8E.142P}
}

@ARTICLE{Caldwell_2016,
       author = {{Caldwell}, Nelson and {Romanowsky}, Aaron J.},
        title = "{Star Clusters in M31. VII. Global Kinematics and Metallicity Subpopulations of the Globular Clusters}",
      journal = {\apj},
         year = 2016,
        month = jun,
       volume = {824},
       number = {1},
          eid = {42},
        pages = {42},
          doi = {10.3847/0004-637X/824/1/42},
archivePrefix = {arXiv},
       eprint = {1603.06947},
 primaryClass = {astro-ph.GA},
       adsurl = {https://ui.adsabs.harvard.edu/abs/2016ApJ...824...42C}
}

@ARTICLE{Dornan_2026,
       author = {{Dornan}, Veronika and {Harris}, William E.},
        title = "{Globular Cluster Systems in Dwarf Galaxies: Catalogs and Comparisons}",
      journal = {\apj},
         year = 2026,
        month = feb,
       volume = {998},
       number = {1},
          eid = {41},
        pages = {41},
          doi = {10.3847/1538-4357/ae2742},
archivePrefix = {arXiv},
       eprint = {2512.08453},
 primaryClass = {astro-ph.GA},
       adsurl = {https://ui.adsabs.harvard.edu/abs/2026ApJ...998...41D}
}

@ARTICLE{Behroozi_2019,
       author = {{Behroozi}, Peter and {Wechsler}, Risa H. and {Hearin}, Andrew P. and {Conroy}, Charlie},
        title = "{UNIVERSEMACHINE: The correlation between galaxy growth and dark matter halo assembly from z = 0-10}",
      journal = {\mnras},
         year = 2019,
        month = sep,
       volume = {488},
       number = {3},
        pages = {3143-3194},
          doi = {10.1093/mnras/stz1182},
archivePrefix = {arXiv},
       eprint = {1806.07893},
 primaryClass = {astro-ph.GA},
       adsurl = {https://ui.adsabs.harvard.edu/abs/2019MNRAS.488.3143B}
}

@ARTICLE{Posti_2019,
       author = {{Posti}, Lorenzo and {Helmi}, Amina},
        title = "{Mass and shape of the Milky Way's dark matter halo with globular clusters from Gaia and Hubble}",
      journal = {\aap},
         year = 2019,
        month = jan,
       volume = {621},
          eid = {A56},
        pages = {A56},
          doi = {10.1051/0004-6361/201833355},
archivePrefix = {arXiv},
       eprint = {1805.01408},
 primaryClass = {astro-ph.GA},
       adsurl = {https://ui.adsabs.harvard.edu/abs/2019A&A...621A..56P}
}

@ARTICLE{Baumgardt_2018,
       author = {{Baumgardt}, H. and {Hilker}, M.},
        title = "{A catalogue of masses, structural parameters, and velocity dispersion profiles of 112 Milky Way globular clusters}",
      journal = {\mnras},
         year = 2018,
        month = aug,
       volume = {478},
       number = {2},
        pages = {1520-1557},
          doi = {10.1093/mnras/sty1057},
archivePrefix = {arXiv},
       eprint = {1804.08359},
 primaryClass = {astro-ph.GA},
       adsurl = {https://ui.adsabs.harvard.edu/abs/2018MNRAS.478.1520B}
}

@misc{BaumgardtDatabase2023,
  author       = {Baumgardt, H. and Sollima, A. and Hilker, M. and et al.},
  title        = {The Galactic Globular Cluster Database},
  year         = {2023},
  howpublished = {\url{https://people.smp.uq.edu.au/HolgerBaumgardt/globular/}},
  note         = {Accessed Month Year}
}

@ARTICLE{Karim_2024,
       author = {{Karim}, Noushin and {Collins}, Michelle L.~M. and {Forbes}, Duncan A. and {Read}, Justin I.},
        title = "{Discovery of Globular Cluster Candidates in the Dwarf Irregular Galaxy IC 2574 Using HST/ACS Imaging}",
      journal = {\mnras},
         year = 2024,
        month = jun,
       volume = {530},
       number = {4},
        pages = {4936-4949},
          doi = {10.1093/mnras/stae611},
archivePrefix = {arXiv},
       eprint = {2402.16955},
 primaryClass = {astro-ph.GA},
       adsurl = {https://ui.adsabs.harvard.edu/abs/2024MNRAS.530.4936K}
}

@ARTICLE{Tudorica_2015,
       author = {{Tudorica}, A. and {Georgiev}, I.~Y. and {Chies-Santos}, A.~L.},
        title = "{Optical-near-IR analysis of globular clusters in the IKN dwarf spheroidal: a complex star formation history}",
      journal = {\aap},
         year = 2015,
        month = sep,
       volume = {581},
          eid = {A84},
        pages = {A84},
          doi = {10.1051/0004-6361/201525615},
archivePrefix = {arXiv},
       eprint = {1506.04155},
 primaryClass = {astro-ph.GA},
       adsurl = {https://ui.adsabs.harvard.edu/abs/2015A&A...581A..84T}
}

@ARTICLE{Fardal_2013,
       author = {{Fardal}, Mark A. and {Weinberg}, Martin D. and {Babul}, Arif and {Irwin}, Mike J. and {Guhathakurta}, Puragra and {Gilbert}, Karoline M. and {Ferguson}, Annette M.~N. and {Ibata}, Rodrigo A. and {Lewis}, Geraint F. and {Tanvir}, Nial R. and {Huxor}, Avon P.},
        title = "{Inferring the Andromeda Galaxy's mass from its giant southern stream with Bayesian simulation sampling}",
      journal = {\mnras},
         year = 2013,
        month = oct,
       volume = {434},
       number = {4},
        pages = {2779-2802},
          doi = {10.1093/mnras/stt1121},
archivePrefix = {arXiv},
       eprint = {1307.3219},
 primaryClass = {astro-ph.CO},
       adsurl = {https://ui.adsabs.harvard.edu/abs/2013MNRAS.434.2779F}
}

@ARTICLE{Rejkuba_2012,
       author = {{Rejkuba}, M.},
        title = "{Globular cluster luminosity function as distance indicator}",
      journal = {\apss},
         year = 2012,
        month = sep,
       volume = {341},
       number = {1},
        pages = {195-206},
          doi = {10.1007/s10509-012-0986-9},
archivePrefix = {arXiv},
       eprint = {1201.3936},
 primaryClass = {astro-ph.CO},
       adsurl = {https://ui.adsabs.harvard.edu/abs/2012Ap&SS.341..195R}
}

@ARTICLE{Zhang_1999,
       author = {{Zhang}, Qing and {Fall}, S. Michael},
        title = "{The Mass Function of Young Star Clusters in the ``Antennae'' Galaxies}",
      journal = {\apjl},
         year = 1999,
        month = dec,
       volume = {527},
       number = {2},
        pages = {L81-L84},
          doi = {10.1086/312412},
archivePrefix = {arXiv},
       eprint = {astro-ph/9911229},
 primaryClass = {astro-ph},
       adsurl = {https://ui.adsabs.harvard.edu/abs/1999ApJ...527L..81Z}
}

@ARTICLE{McCrady_2007,
       author = {{McCrady}, Nate and {Graham}, James R.},
        title = "{Super Star Cluster Velocity Dispersions and Virial Masses in the M82 Nuclear Starburst}",
      journal = {\apj},
         year = 2007,
        month = jul,
       volume = {663},
       number = {2},
        pages = {844-856},
          doi = {10.1086/518357},
archivePrefix = {arXiv},
       eprint = {0704.0478},
 primaryClass = {astro-ph},
       adsurl = {https://ui.adsabs.harvard.edu/abs/2007ApJ...663..844M}
}

@ARTICLE{Vesperini_2001,
       author = {{Vesperini}, E.},
        title = "{Evolution of globular cluster systems in elliptical galaxies - II. Power-law initial mass function}",
      journal = {\mnras},
         year = 2001,
        month = apr,
       volume = {322},
       number = {2},
        pages = {247-256},
          doi = {10.1046/j.1365-8711.2001.04072.x},
archivePrefix = {arXiv},
       eprint = {astro-ph/0010111},
 primaryClass = {astro-ph},
       adsurl = {https://ui.adsabs.harvard.edu/abs/2001MNRAS.322..247V}
}

@ARTICLE{McLaughlin_1994,
       author = {{McLaughlin}, Dean E.},
        title = "{An Analytical Study of the Globular-Cluster Luminosity Function}",
      journal = {\pasp},
         year = 1994,
        month = jan,
       volume = {106},
        pages = {47},
          doi = {10.1086/133340},
       adsurl = {https://ui.adsabs.harvard.edu/abs/1994PASP..106...47M}
}

@ARTICLE{McLaughlin_1996,
       author = {{McLaughlin}, Dean E. and {Pudritz}, Ralph E.},
        title = "{The Formation of Globular Cluster Systems. I. The Luminosity Function}",
      journal = {\apj},
         year = 1996,
        month = feb,
       volume = {457},
        pages = {578},
          doi = {10.1086/176754},
       adsurl = {https://ui.adsabs.harvard.edu/abs/1996ApJ...457..578M}
}

@ARTICLE{Parmentier_2007,
       author = {{Parmentier}, Genevi{\`e}ve and {Gilmore}, Gerard},
        title = "{The origin of the Gaussian initial mass function of old globular cluster systems}",
      journal = {\mnras},
         year = 2007,
        month = may,
       volume = {377},
       number = {1},
        pages = {352-372},
          doi = {10.1111/j.1365-2966.2007.11611.x},
archivePrefix = {arXiv},
       eprint = {astro-ph/0702258},
 primaryClass = {astro-ph},
       adsurl = {https://ui.adsabs.harvard.edu/abs/2007MNRAS.377..352P}
}

@ARTICLE{Gnedin_1997,
       author = {{Gnedin}, Oleg Y. and {Ostriker}, Jeremiah P.},
        title = "{Destruction of the Galactic Globular Cluster System}",
      journal = {\apj},
         year = 1997,
        month = jan,
       volume = {474},
       number = {1},
        pages = {223-255},
          doi = {10.1086/303441},
archivePrefix = {arXiv},
       eprint = {astro-ph/9603042},
 primaryClass = {astro-ph},
       adsurl = {https://ui.adsabs.harvard.edu/abs/1997ApJ...474..223G}
}

@ARTICLE{Claeyssens2026,
       author = {{Claeyssens}, A. and {Adamo}, A. and {Kokorev}, V. and {Furtak}, L. and {Richard}, J. and {Beauchesne}, B. and {Dessauges-Zavadsky}, M. and {Atek}, H. and {Chisholm}, J. and {Endsley}, R. and {Fujimoto}, S. and {Korber}, D. and {Pan}, R. and {Saldana-Lopez}, A. and {Schaerer}, D.},
        title = "{A first GLIMPSE into star-cluster populations across cosmic time}",
      journal = {\aap},
         year = 2026,
        month = oct,
       volume = {714},
          eid = {A148},
        pages = {A148},
          doi = {10.1051/0004-6361/202659135},
       adsurl = {https://ui.adsabs.harvard.edu/abs/2026A&A...714A.148C}
}

@ARTICLE{Vanzella_2026,
       author = {{Vanzella}, E. and {Messa}, M. and {Adamo}, A. and {Loiacono}, F. and {Oguri}, M. and {Sharon}, K. and {Bradley}, L.~D. and {Bergamini}, P. and {Meneghetti}, M. and {Claeyssens}, A. and {Welch}, B. and {Brada{\v{c}}}, M. and {Zanella}, A. and {Bolamperti}, A. and {Calura}, F. and {Hsiao}, T.~Y.-Y. and {Zackrisson}, E. and {Ricotti}, M. and {Christensen}, L. and {Diego}, J.~M. and {Bauer}, F.~E. and {Xu}, X. and {Fujimoto}, S. and {Grillo}, C. and {Lombardi}, M. and {Rosati}, P. and {Resseguier}, T. and {Zitrin}, A. and {Bik}, A. and {Richard}, J. and {Abdurro'uf} and {Bhatawdekar}, R. and {Coe}, D. and {Frye}, B. and {Inoue}, A.~K. and {Jimenez-Teja}, Y. and {Norman}, C. and {Rigby}, J.~R. and {Trenti}, M. and {Hashimoto}, T.},
        title = "{The z = 9.625 Cosmic Gems galaxy was a compact ``blue monster'' propelled by massive star clusters}",
      journal = {\aap},
         year = 2026,
        month = jan,
       volume = {705},
          eid = {A171},
        pages = {A171},
          doi = {10.1051/0004-6361/202556570},
archivePrefix = {arXiv},
       eprint = {2507.18699},
 primaryClass = {astro-ph.GA},
       adsurl = {https://ui.adsabs.harvard.edu/abs/2026A&A...705A.171V}
}

@ARTICLE{Jordan2009,
       author = {{Jord{\'a}n}, Andr{\'e}s and {Peng}, Eric W. and {Blakeslee}, John P. and {C{\^o}t{\'e}}, Patrick and {Eyheramendy}, Susana and {Ferrarese}, Laura and {Mei}, Simona and {Tonry}, John L. and {West}, Michael J.},
        title = "{The ACS Virgo Cluster Survey XVI. Selection Procedure and Catalogs of Globular Cluster Candidates}",
      journal = {\apjs},
         year = 2009,
        month = jan,
       volume = {180},
       number = {1},
        pages = {54-66},
          doi = {10.1088/0067-0049/180/1/54},
       adsurl = {https://ui.adsabs.harvard.edu/abs/2009ApJS..180...54J}
}

@ARTICLE{Bell2003,
       author = {{Bell}, Eric F. and {McIntosh}, Daniel H. and {Katz}, Neal and {Weinberg}, Martin D.},
        title = "{The Optical and Near-Infrared Properties of Galaxies. I. Luminosity and Stellar Mass Functions}",
      journal = {\apjs},
         year = 2003,
        month = dec,
       volume = {149},
       number = {2},
        pages = {289-312},
          doi = {10.1086/378847},
archivePrefix = {arXiv},
       eprint = {astro-ph/0302543},
 primaryClass = {astro-ph},
       adsurl = {https://ui.adsabs.harvard.edu/abs/2003ApJS..149..289B}
}

@ARTICLE{Adamo2024,
       author = {{Adamo}, Angela and {Bradley}, Larry D. and {Vanzella}, Eros and {Claeyssens}, Ad{\'e}la{\"\i}de and {Welch}, Brian and {Diego}, Jose M. and {Mahler}, Guillaume and {Oguri}, Masamune and {Sharon}, Keren and {Abdurro'uf} and {Hsiao}, Tiger Yu-Yang and {Xu}, Xinfeng and {Messa}, Matteo and {Lassen}, Augusto E. and {Zackrisson}, Erik and {Brammer}, Gabriel and {Coe}, Dan and {Kokorev}, Vasily and {Ricotti}, Massimo and {Zitrin}, Adi and {Fujimoto}, Seiji and {Inoue}, Akio K. and {Resseguier}, Tom and {Rigby}, Jane R. and {Jim{\'e}nez-Teja}, Yolanda and {Windhorst}, Rogier A. and {Hashimoto}, Takuya and {Tamura}, Yoichi},
        title = "{Bound star clusters observed in a lensed galaxy 460 Myr after the Big Bang}",
      journal = {\nat},
         year = 2024,
        month = aug,
       volume = {632},
       number = {8025},
        pages = {513-516},
          doi = {10.1038/s41586-024-07703-7},
archivePrefix = {arXiv},
       eprint = {2401.03224},
 primaryClass = {astro-ph.GA},
       adsurl = {https://ui.adsabs.harvard.edu/abs/2024Natur.632..513A}
}

@ARTICLE{Garcia2023,
       author = {{Garcia}, Fred Angelo Batan and {Ricotti}, Massimo and {Sugimura}, Kazuyuki and {Park}, Jongwon},
        title = "{Star cluster formation and survival in the first galaxies}",
      journal = {\mnras},
         year = 2023,
        month = jun,
       volume = {522},
       number = {2},
        pages = {2495-2515},
          doi = {10.1093/mnras/stad1092},
archivePrefix = {arXiv},
       eprint = {2212.13946},
 primaryClass = {astro-ph.GA},
       adsurl = {https://ui.adsabs.harvard.edu/abs/2023MNRAS.522.2495G}
}

@ARTICLE{Andersson2024,
       author = {{Andersson}, Eric P. and {Mac Low}, Mordecai-Mark and {Agertz}, Oscar and {Renaud}, Florent and {Li}, Hui},
        title = "{Pre-supernova feedback sets the star cluster mass function to a power law and reduces the cluster formation efficiency}",
      journal = {\aap},
         year = 2024,
        month = jan,
       volume = {681},
          eid = {A28},
        pages = {A28},
          doi = {10.1051/0004-6361/202347792},
archivePrefix = {arXiv},
       eprint = {2308.12363},
 primaryClass = {astro-ph.GA},
       adsurl = {https://ui.adsabs.harvard.edu/abs/2024A&A...681A..28A}
}

@ARTICLE{Moreno-Hilario2024,
       author = {{Moreno-Hilario}, Elizabeth and {Martinez-Medina}, Luis A. and {Li}, Hui and {Souza}, Stefano O. and {P{\'e}rez-Villegas}, Angeles},
        title = "{The influence of globular cluster evolution on the specific frequency in dwarf galaxies}",
      journal = {\mnras},
         year = 2024,
        month = jan,
       volume = {527},
       number = {2},
        pages = {2765-2780},
          doi = {10.1093/mnras/stad3306},
archivePrefix = {arXiv},
       eprint = {2310.17396},
 primaryClass = {astro-ph.GA},
       adsurl = {https://ui.adsabs.harvard.edu/abs/2024MNRAS.527.2765M}
}

@ARTICLE{Kimm2016,
       author = {{Kimm}, Taysun and {Cen}, Renyue and {Rosdahl}, Joakim and {Yi}, Sukyoung K.},
        title = "{Formation of Globular Clusters in Atomic-cooling Halos Via Rapid Gas Condensation and Fragmentation during the Epoch of Reionization}",
      journal = {\apj},
         year = 2016,
        month = may,
       volume = {823},
       number = {1},
          eid = {52},
        pages = {52},
          doi = {10.3847/0004-637X/823/1/52},
archivePrefix = {arXiv},
       eprint = {1510.05671},
 primaryClass = {astro-ph.GA},
       adsurl = {https://ui.adsabs.harvard.edu/abs/2016ApJ...823...52K}
}

@ARTICLE{Rieder2022,
       author = {{Rieder}, Steven and {Dobbs}, Clare and {Bending}, Thomas and {Liow}, Kong You and {Wurster}, James},
        title = "{The formation and early evolution of embedded star clusters in spiral galaxies}",
      journal = {\mnras},
         year = 2022,
        month = feb,
       volume = {509},
       number = {4},
        pages = {6155-6168},
          doi = {10.1093/mnras/stab3425},
archivePrefix = {arXiv},
       eprint = {2111.09720},
 primaryClass = {astro-ph.GA},
       adsurl = {https://ui.adsabs.harvard.edu/abs/2022MNRAS.509.6155R}
}

@ARTICLE{Chen2023,
       author = {{Chen}, Yingtian and {Gnedin}, Oleg Y.},
        title = "{Formation of globular clusters in dwarf galaxies of the Local Group}",
      journal = {\mnras},
         year = 2023,
        month = jul,
       volume = {522},
       number = {4},
        pages = {5638-5653},
          doi = {10.1093/mnras/stad1328},
archivePrefix = {arXiv},
       eprint = {2301.08218},
 primaryClass = {astro-ph.GA},
       adsurl = {https://ui.adsabs.harvard.edu/abs/2023MNRAS.522.5638C}
}

@ARTICLE{Rodriguez2023,
       author = {{Rodriguez}, Carl L. and {Hafen}, Zachary and {Grudi{\'c}}, Michael Y. and {Lamberts}, Astrid and {Sharma}, Kuldeep and {Faucher-Gigu{\`e}re}, Claude-Andr{\'e} and {Wetzel}, Andrew},
        title = "{Great balls of FIRE II: The evolution and destruction of star clusters across cosmic time in a Milky Way-mass galaxy}",
      journal = {\mnras},
         year = 2023,
        month = may,
       volume = {521},
       number = {1},
        pages = {124-147},
          doi = {10.1093/mnras/stad578},
archivePrefix = {arXiv},
       eprint = {2203.16547},
 primaryClass = {astro-ph.GA},
       adsurl = {https://ui.adsabs.harvard.edu/abs/2023MNRAS.521..124R}
}

@ARTICLE{Forbes2018,
       author = {{Forbes}, Duncan A. and {Read}, Justin I. and {Gieles}, Mark and {Collins}, Michelle L.~M.},
        title = "{Extending the globular cluster system-halo mass relation to the lowest galaxy masses}",
      journal = {\mnras},
         year = 2018,
        month = dec,
       volume = {481},
       number = {4},
        pages = {5592-5605},
          doi = {10.1093/mnras/sty2584},
archivePrefix = {arXiv},
       eprint = {1809.07831},
 primaryClass = {astro-ph.GA},
       adsurl = {https://ui.adsabs.harvard.edu/abs/2018MNRAS.481.5592F}
}

@ARTICLE{Fall2001,
       author = {{Fall}, S. Michael and {Zhang}, Qing},
        title = "{Dynamical Evolution of the Mass Function of Globular Star Clusters}",
      journal = {\apj},
         year = 2001,
        month = nov,
       volume = {561},
       number = {2},
        pages = {751-765},
          doi = {10.1086/323358},
archivePrefix = {arXiv},
       eprint = {astro-ph/0107298},
 primaryClass = {astro-ph},
       adsurl = {https://ui.adsabs.harvard.edu/abs/2001ApJ...561..751F}
}

@ARTICLE{Banerjee2020,
       author = {{Banerjee}, S. and {Belczynski}, K. and {Fryer}, C.~L. and {Berczik}, P. and {Hurley}, J.~R. and {Spurzem}, R. and {Wang}, L.},
        title = "{BSE versus StarTrack: Implementations of new wind, remnant-formation, and natal-kick schemes in NBODY7 and their astrophysical consequences}",
      journal = {\aap},
         year = 2020,
        month = jul,
       volume = {639},
          eid = {A41},
        pages = {A41},
          doi = {10.1051/0004-6361/201935332},
archivePrefix = {arXiv},
       eprint = {1902.07718},
 primaryClass = {astro-ph.SR},
       adsurl = {https://ui.adsabs.harvard.edu/abs/2020A&A...639A..41B}
}

@ARTICLE{Wang2015,
       author = {{Wang}, Long and {Spurzem}, Rainer and {Aarseth}, Sverre and {Nitadori}, Keigo and {Berczik}, Peter and {Kouwenhoven}, M.~B.~N. and {Naab}, Thorsten},
        title = "{NBODY6++GPU: ready for the gravitational million-body problem}",
      journal = {\mnras},
         year = 2015,
        month = jul,
       volume = {450},
       number = {4},
        pages = {4070-4080},
          doi = {10.1093/mnras/stv817},
archivePrefix = {arXiv},
       eprint = {1504.03687},
 primaryClass = {astro-ph.IM},
       adsurl = {https://ui.adsabs.harvard.edu/abs/2015MNRAS.450.4070W}
}

@ARTICLE{Banerjee2011,
       author = {{Banerjee}, Sambaran and {Kroupa}, Pavel},
        title = "{A New Type of Compact Stellar Population: Dark Star Clusters}",
      journal = {\apjl},
         year = 2011,
        month = nov,
       volume = {741},
       number = {1},
          eid = {L12},
        pages = {L12},
          doi = {10.1088/2041-8205/741/1/L12},
archivePrefix = {arXiv},
       eprint = {1110.4103},
 primaryClass = {astro-ph.GA},
       adsurl = {https://ui.adsabs.harvard.edu/abs/2011ApJ...741L..12B}
}

@ARTICLE{Giersz2019,
       author = {{Giersz}, M. and {Askar}, A. and {Wang}, L. and {Hypki}, A. and {Leveque}, A. and {Spurzem}, R.},
        title = "{MOCCA survey data base- I. Dissolution of tidally filling star clusters harbouring black hole subsystems}",
      journal = {\mnras},
         year = 2019,
        month = aug,
       volume = {487},
       number = {2},
        pages = {2412-2423},
          doi = {10.1093/mnras/stz1460},
archivePrefix = {arXiv},
       eprint = {1904.01227},
 primaryClass = {astro-ph.GA},
       adsurl = {https://ui.adsabs.harvard.edu/abs/2019MNRAS.487.2412G}
}

@ARTICLE{Kremer2020,
       author = {{Kremer}, Kyle and {Ye}, Claire S. and {Rui}, Nicholas Z. and {Weatherford}, Newlin C. and {Chatterjee}, Sourav and {Fragione}, Giacomo and {Rodriguez}, Carl L. and {Spera}, Mario and {Rasio}, Frederic A.},
        title = "{Modeling Dense Star Clusters in the Milky Way and Beyond with the CMC Cluster Catalog}",
      journal = {\apjs},
         year = 2020,
        month = apr,
       volume = {247},
       number = {2},
          eid = {48},
        pages = {48},
          doi = {10.3847/1538-4365/ab7919},
archivePrefix = {arXiv},
       eprint = {1911.00018},
 primaryClass = {astro-ph.HE},
       adsurl = {https://ui.adsabs.harvard.edu/abs/2020ApJS..247...48K}
}

@ARTICLE{Wang2020,
       author = {{Wang}, Long},
        title = "{The survival of star clusters with black hole subsystems}",
      journal = {\mnras},
         year = 2020,
        month = jan,
       volume = {491},
       number = {2},
        pages = {2413-2423},
          doi = {10.1093/mnras/stz3179},
archivePrefix = {arXiv},
       eprint = {1911.05077},
 primaryClass = {astro-ph.GA},
       adsurl = {https://ui.adsabs.harvard.edu/abs/2020MNRAS.491.2413W}
}

@ARTICLE{Gieles2011,
       author = {{Gieles}, Mark and {Heggie}, Douglas C. and {Zhao}, Hongsheng},
        title = "{The life cycle of star clusters in a tidal field}",
      journal = {\mnras},
         year = 2011,
        month = jun,
       volume = {413},
       number = {4},
        pages = {2509-2524},
          doi = {10.1111/j.1365-2966.2011.18320.x},
archivePrefix = {arXiv},
       eprint = {1101.1821},
 primaryClass = {astro-ph.GA},
       adsurl = {https://ui.adsabs.harvard.edu/abs/2011MNRAS.413.2509G}
}
\bibliographystyle{aasjournalv7}



\end{document}